\documentclass[aps,prx,twocolumn,superscriptaddress]{revtex4-2}

\usepackage{graphicx}% Include figure files
\usepackage{dcolumn}% Align table columns on decimal point
\usepackage{bm}% bold math
\usepackage{braket}
\usepackage{amsfonts}
\usepackage{amsmath}
\usepackage{hyperref}% add hypertext capabilities
\hypersetup{
	colorlinks=true,
	linkcolor=blue,
	filecolor=magenta,      
	urlcolor=blue,
	citecolor=blue
}

\graphicspath{ {./Figures/exported_images/PDF/} }
\newcommand{\ketbra}[2]{|#1\rangle\langle #2|}
\begin{document}

\preprint{}

\title{Single-shot number-resolved detection of microwave photons with error mitigation}% Force line breaks with \\

\author{Jacob C. Curtis}
\email{jacob.curtis@yale.edu}
\author{Connor T. Hann}
\author{Salvatore S. Elder}
\author{Christopher S. Wang}
\author{Luigi Frunzio}
\affiliation{Departments of Physics and Applied Physics, Yale University, New Haven, Connecticut 06511, USA}
\affiliation{Yale Quantum Institute, Yale University, New Haven, Connecticut 06511, USA}
\author{Liang Jiang}
\affiliation{ Pritzker School of Molecular Engineering, University of Chicago, Chicago, Illinois 60637, USA}
\author{Robert J. Schoelkopf}
 \email{robert.schoelkopf@yale.edu}
\affiliation{Departments of Physics and Applied Physics, Yale University, New Haven, Connecticut 06511, USA}
\affiliation{Yale Quantum Institute, Yale University, New Haven, Connecticut 06511, USA}

\date{\today}
\begin{abstract}
Single-photon detectors are ubiquitous and integral components of photonic quantum cryptography, communication, and computation. Many applications, however, require not only detecting the presence of any photons, but distinguishing the number present with a single shot. Here, we implement a single-shot, high-fidelity photon number-resolving detector of up to 15 microwave photons in a cavity-qubit circuit QED platform. This detector functions by measuring a series of generalized parity operators which make up the bits in the binary decomposition of the photon number. Our protocol consists of successive, independent measurements of each bit by entangling the ancilla with the cavity, then reading out and resetting the ancilla. Photon loss and ancilla readout errors can flip one or more bits, causing nontrivial errors in the outcome, but these errors have a traceable form which can be captured in a simple hidden Markov model. Relying on the independence of each bit measurement, we mitigate biases in ensembles of measurements, showing good agreement with the predictions of the model. The mitigation improves the average total variation distance error of Fock states from $13.5\%$ to $1.1\%$. We also show that the mitigation is efficiently scalable to an $M$-mode system provided that the errors are independent and sufficiently small. Our work motivates the development of new algorithms that utilize single-shot, high-fidelity PNR detectors.
\end{abstract}

\maketitle

\begin{figure*}
	\includegraphics[scale=1]{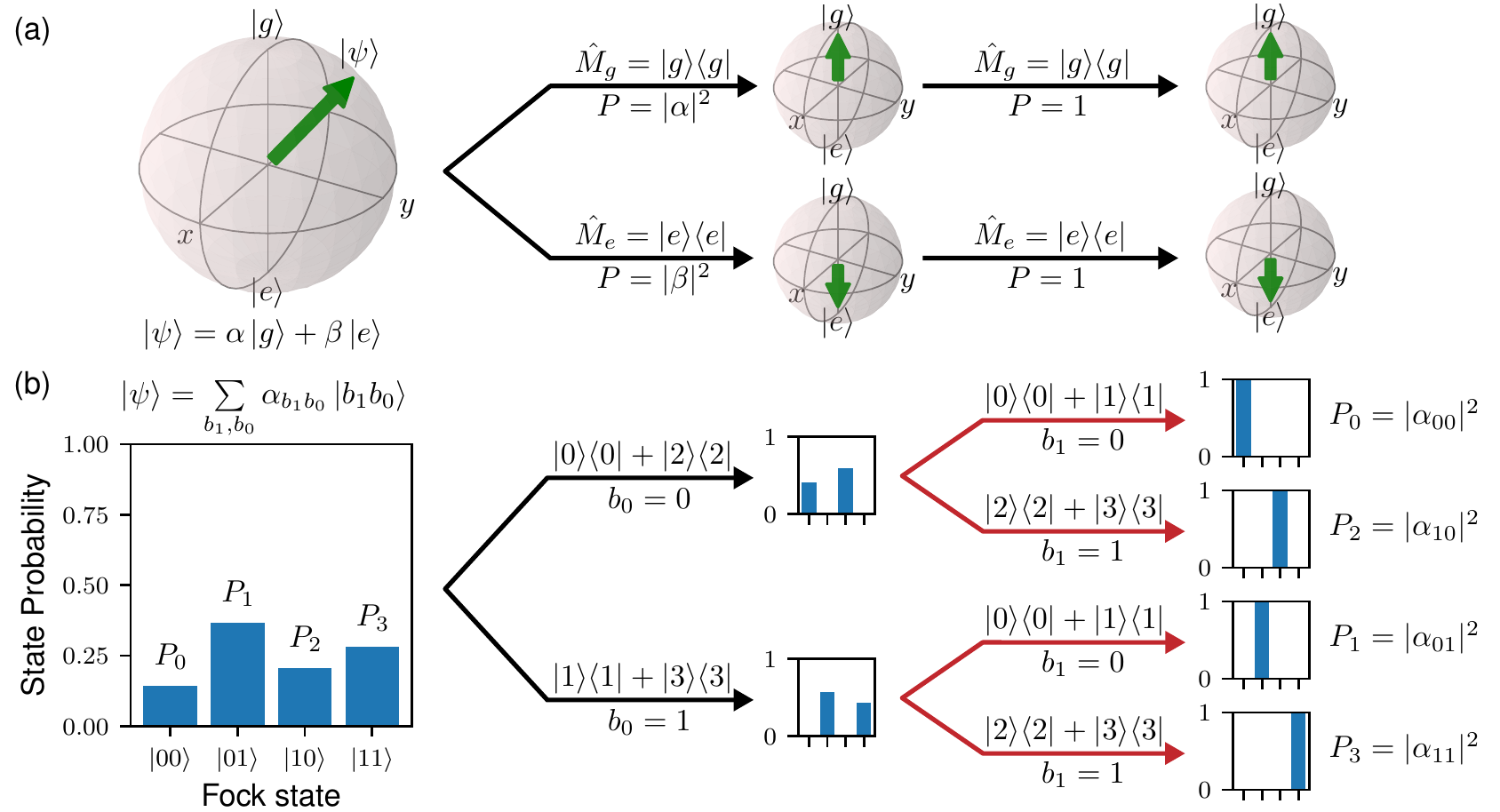}
	\caption{\label{fig1} Projective, binary-valued measurements. (a) Measuring a qubit in its energy eigenbasis (the basis of $\sigma_z$ eigenstates) projects the qubit into its ground state $\ket{g}$ with probability $|\alpha|^2$ or its excited state $\ket{e}$ with probability $|\beta|^2$. The measurement operators $\hat{M}_g$ and $\hat{M}_e$ commute with the Hamiltonian at all times. As a result, the shared eigenstates are not perturbed by additional measurements or time evolution, making the measurement QND. (b) Here, Fock states are represented with the first two bits (parity and super-parity) of their binary decomposition $\ket{n}=\ket{b_1(n)b_0(n)}$ for $N<4$. An initial measurement of parity (black arrows labeled by measurement outcome and corresponding measurement operator) projects $\ket{\psi}$ into the even or odd subspace, and renormalizes the remaining state probabilities. A subsequent measurement of super-parity (red arrows), further projects the state fully into a single Fock state $\ket{n}$. The detected photon number is computed using a record of each measurement.}
\end{figure*}

\section{Introduction}
Many quantum information processing protocols that utilize photonic platforms require devices that detect single photons. These single photon detectors (SPDs) perform essential tasks such as measuring the output of a photonic register or heralding in probabilistic quantum gates. Uses of SPDs include quantum key distribution \cite{lo_measurement-device-independent_2012, yin_measurement-device-independent_2016, liao_satellite--ground_2017}, linear optical quantum computing (LOQC) \cite{knill_scheme_2001, kok_linear_2007}, quantum communication \cite{duan_long-distance_2001, ursin_entanglement-based_2007, kimble_quantum_2008, northup_quantum_2014, hu_experimental_2016, zhang_quantum_2017, dou_broadband_2018}, and photonic quantum simulations \cite{spring_boson_2013, tillmann_experimental_2013, sparrow_simulating_2018, bentivegna_experimental_2015}. Highly efficient SPDs with low dark count rates based on avalanche photodiodes and, more recently, superconducting nanowire single-photon detectors \cite{eisaman_invited_2011,marsili_detecting_2013, esmaeil_zadeh_single-photon_2017} have been developed and satisfy the demands of these optical applications. 

The information processing capacity of intrinsically bosonic optical modes, however, can be much greater by executing protocols that manipulate multiphoton states. In these cases, SPDs are insufficient for distinguishing between multiple photons and photon number-resolving (PNR) detectors are required instead. Proposals to enhance quantum communication \cite{simon_quantum_2007} and key distribution \cite{cattaneo_hybrid_2018} protocols, perform teleported gates \cite{knill_scheme_2001}, and extend conventional boson sampling \cite{spring_boson_2013, huh_boson_2015,hamilton_gaussian_2017,clements_approximating_2018,  kruse_detailed_2019} all require PNR detectors. Building an optical PNR detector has proven to be a difficult task. There is a variety of promising approaches relying on highly efficient SPDs in multiplexing or arraying schemes \cite{banaszek_photon_2003, fitch_photon-number_2003, achilles_photon-number-resolving_2004, divochiy_superconducting_2008, mattioli_photon-counting_2016, tao_high_2019} and transition edge sensors \cite{kardynal_avalanchephotodiode-based_2008,gerrits_-chip_2011, calkins_high_2013}. These recent advances have vastly improved performance, but the limited fidelity and resolvable photon number of optical PNR detectors hamper multiphoton experiments \cite{clements_approximating_2018}.

The detection of single microwave photons is less established and more challenging, due to their lower energy and higher thermal background levels. The dispersive interaction between a photon and an atom or qubit enables a wide range of measurement capabilities not possible in the optical domain. In cavity QED, the dispersive interaction with Rydberg atoms enabled the observation of single photon jumps \cite{gleyzes_quantum_2007}. In circuit QED (cQED) systems, Josephson junctions coupled to microwave cavities generate the dispersive interaction essential for measuring single microwave photons \cite{schuster_resolving_2007, johnson_quantum_2010}. Additionally, the dispersive interaction has been used to create \cite{houck_generating_2007, hofheinz_generation_2008} and manipulate \cite{heeres_cavity_2015, heeres_implementing_2017} these photons in a wide variety of ways. cQED analogs of optical photodetectors such as the Josephson photomultiplier \cite{chen_microwave_2011, govia_high-fidelity_2014, opremcak_measurement_2018} have also been developed.

Introducing high-performance PNR detectors to the microwave regime could greatly enhance the prospects for performing boson sampling and other photonic quantum information processing techniques with cQED systems. Early quantum non-demolition (QND), single-shot, number-resolving measurements of microwave photons used flying Rydberg atoms to iteratively update the experimentalists' knowledge of the photon number distribution \cite{guerlin_progressive_2007}. Previous efforts in cQED to develop PNR measurements have fallen short of combining single-shot and QND capabilities. Spectral density analysis \cite{hofheinz_generation_2008} and state interrogation methods \cite{schuster_resolving_2007} are QND, but require many shots of the experiment to build up time or spectroscopic traces. Parity measurements have been used to count photon jumps \cite{sun_tracking_2014}, but, while single-shot, the full PNR measurement is not QND. More recently, frequency multiplexing \cite{essig_multiplexed_2020} and multiple feedforward measurements \cite{dassonneville_number-resolved_2020} have been used to make QND, single-shot PNR measurements, but suffer from limited fidelity.

In this paper, we describe a measurement protocol that implements a high-fidelity, single-shot PNR detector of microwave photons. The single-shot nature of this measurement is essential to sample from the exponentially large Hilbert space inherent to experiments with multiple bosonic modes. We previously introduced this protocol to efficiently sample from the probability distribution at the output of a multiphoton bosonic quantum simulator, reducing the required number of measurements by a factor of 256, the size of the Hilbert space \cite{wang_efficient_2020}. Errors in the storage and ancilla modes caused non-trivial bias in the output, but these errors are well understood and can be modeled with a simple hidden Markov model. We calibrate and use this model to improve the fidelity of the measurement by an order of magnitude using error mitigation methods, reducing the measurement infidelity to a few percent.

This paper is structured as follows. First in Sec.~\ref{QND single-shot}, we discuss QND measurements of multiple observables with a single prepared state. We verify measurements of parity and its generalizations satisfy the conditions required to be QND. Next, we show how to construct a PNR detector of microwave photons from successive QND measurements of parity and its generalizations. This is accomplished by representing the photon number in its binary decomposition. In Sec.~\ref{sec_3}, we introduce the error syndromes of our system along with a hidden Markov model (HMM) that parametrizes the errors. After calibrating the error rates, we use the model to mitigate the errors on an ensemble of states using deconvolution methods. Finally, we show that our error deconvolution protocol is scalable in the number of modes and maintains computational efficiency.

\section{Background}
\label{QND single-shot}
Optical photon detectors such as photomultiplier tubes, avalanche photodiodes, and superconducting nanowire SPDs function by converting incoming photons into electrical charges which are then amplified to produce a detectable signal \cite{eisaman_invited_2011}. This process is single-shot, but is inherently destructive as the incoming photons are consumed by the detector. Standard techniques in cQED for measuring single photons are QND \cite{johnson_quantum_2010, sun_tracking_2014}, allowing for additional processing of the state. Our protocol for PNR detection utilizes this ability to perform multiple measurements that collectively form a number-resolving measurement. In this section, we generalize the notions of QND and single-shot to a multi-level system with multiple measurements.

\subsection{Multi-level QND measurements}
To realize a single-shot PNR detector, we need to extend the concept of QND measurements to the case of a multi-level system and multiple measurements. Not all observables can be measured in a QND way; they must satisfy two conditions \cite{haroche_exploring_nodate}. The first QND condition is that the observable $\hat{\mathcal{A}}$ commutes with the system Hamiltonian $\hat{H}$ to ensure that a repeated measurement of it yields the same result at any subsequent time (in the absence of external perturbations, such as photon loss). The second condition is that $\hat{\mathcal{A}}$ commutes with the measurement Hamiltonian $\hat{H}_M$, which contains additional terms that couple to a meter. This condition protects the state from additional evolution induced by the measurement process. QND measurements implement measurement operators $\hat{M}_{\lambda}$ for each outcome $\lambda$, which project the system into the eigenstate labeled by its eigenvalue $\lambda$.

Let us first consider the usual case of QND measurements of a qubit or two-level system initialized in an arbitrary state $\ket{\psi}=\alpha\ket{g}+\beta\ket{e}$. A measurement of the system in this basis, with measurement operators $\hat{M}_g=\ketbra{g}{g}$ and $\hat{M}_e=\ketbra{e}{e}$, projects $\ket{\psi}$ into $\ket{g}$ with probability $|\alpha|^2$ and $\ket{e}$ with probability $|\beta|^2$. The outcome of this measurement tells us exactly the state of the qubit following the measurement. If the measurement of the observable is QND, then repeated measurements at any subsequent time will yield the same result, as seen in Fig.~\ref{fig1} (a).

How can we extend this concept to perform a QND measurement on a multilevel system? For instance, consider a four-level subspace of a bosonic mode $s$ with operators $\hat{s},\hat{s}^\dagger$. Two bits of classical information must be obtained to determine the state of this system. This can be accomplished with two QND measurements, whose composition projects the system into a single level. The first QND measurement halves the Hilbert space, and projects the state into one of its two subspaces, extracting one bit of classical information. A second measurement further halves this subspace, extracting the second bit and projecting the state into a single level of $s$. This is only possible if the two operators commute, else the second measurement would project into a superposition of states from the two subspaces of the first measurement, spoiling the first bit of information.

For example, the binary decomposition of the number of excitations in $s$ contains two bits identifying the state of the four level system discussed above. Parity $\hat{\mathcal{P}}^0$ divides the space into even and odd subspaces with measurement operators $\hat{B}^{(0)}_{b_0=0}=\ketbra{0}{0}+\ketbra{2}{2}$ and $\hat{B}^{(0)}_{b_0=1}=\ketbra{1}{1}+\ketbra{3}{3}$. This measurement projects the system into the state corresponding to measurement outcome $b_0$, which is the least significant bit in the binary decomposition of the excitation number of $s$. Measurements of super-parity $\hat{\mathcal{P}}^1$, which commutes with $\hat{\mathcal{P}}^0$, measure the second-least-significant bit $b_1$ with measurement operators $\hat{B}^{(1)}_{b_1=0}=\ketbra{0}{0}+\ketbra{1}{1}$ and $\hat{B}^{(1)}_{b_1=1}=\ketbra{2}{2}+\ketbra{3}{3}$. Measured in succession, these two observables project the system into a single Fock state labeled by its binary decomposition $\ket{b_1b_0}$, as shown in Fig.~\ref{fig1} (b). These generalized parity measurements form the basis of our approach to realize a QND, PNR detector of microwave photons in a cQED system.

\subsection{Multi-level single-shot measurements}
\label{single shot}
As we discuss in the previous section, measuring a two-level or multi-level system requires extracting one or more bits of information. For these measurements to be considered single-shot, they must accurately measure a significant majority of their bits in a single shot. When measuring a qubit, this means that one gains close to one bit of information about the state per measurement, per initial state subjected to that measurement. With a correspondingly high measurement efficiency and a low dark count rate, a qubit measurement then extracts enough of the available information that we can measure it with precision close to the shot noise limit imposed by the ensemble size.

We call a multi-level measurement single-shot if a significant fraction of the bits are faithfully extracted per shot. For example, choosing $3/4$ as this fraction corresponds to extracting more than three bits of information per shot for a PNR detector with $N=16$ levels. To determine $P(n)$ of a multi-level system, we must sample many times from an ensemble of states. If the PNR measurement that acquires each sample meets this threshold, the error in the sampled $P(n)$ is limited by the shot noise of the number of samples. This capability is particularly crucial to extract joint probability distributions from $M$ $N$-level systems without individually checking each of the $N^M$ states. We discuss the single-shot character of our detector in Sec.~\ref{model calibration}.

\section{The bitwise measurement}
\subsection{Measuring binary-valued cavity observables}
Our PNR detector uses a qubit ancilla to measure binary-valued observables of a bosonic storage mode via the dispersive interaction. These observables $\hat{\mathcal{O}}$ divide the Hilbert space into exactly two subspaces with distinct eigenvalues $\lambda_1,\lambda_2$. We can imagine a gate that entangles the two ancilla states with the eigenspaces $E_{\lambda_1},E_{\lambda_2}$. A readout of the ancilla would then project the storage mode into the corresponding eigenspace. This two-step measurement is QND only if the QND conditions are satisfied for all steps in the measurement. For a measurement of $\hat{\mathcal{O}}$ to satisfy the first QND condition, it must be diagonal in the energy eigenbasis.

We next verify that the second QND condition is satisfied. Our three-mode cQED system consisting of a readout mode, a storage mode, and an ancilla has operators $\hat{r},\hat{r}^\dagger$, $\hat{s},\hat{s}^\dagger$, and $\hat{\sigma}_z$, respectively. The composite system Hamiltonian is
\begin{equation}
\hat{H}/\hbar=\frac{\omega_{\sigma}}{2}\hat{\sigma}_z+ \omega_s \hat{s}^\dagger\hat{s}+\omega_r \hat{r}^\dagger\hat{r}-\frac{\chi_{s\sigma}}{2}\hat{s}^\dagger\hat{s}\hat{\sigma}_z-\frac{\chi_{r\sigma}}{2}\hat{r}^\dagger\hat{r}\hat{\sigma}_z.
\end{equation}
The device is the same as used in Refs. \cite{elder_high-fidelity_2020,reinhold_error-corrected_2020} with similar parameters, unless otherwise noted. We first apply an ancilla-storage entangling drive, which produces the mapping Hamiltonian $\hat{H}_{\text{map}}/\hbar=\hat{H}/\hbar+\Omega(t)\hat{\sigma}_x$. For observables $\hat{\mathcal{O}}$, frequency-selective drives $\Omega(t)$ can implement the requisite entangling operation by exciting the ancilla conditioned on a Fock state. The second step, ancilla readout, also utilizes a dispersive interaction, realized with the Hamiltonian $\hat{H}_{\text{readout}}/\hbar=\hat{H}/\hbar+\epsilon(t)\hat{r}+\epsilon^*(t)\hat{r}^\dagger$. This readout completes the measurement of $\hat{\mathcal{O}}$. Both the mapping and readout Hamiltonians commute with the diagonal $\hat{\mathcal{O}}$, and satisfy the second QND condition. These drives can induce dephasing \cite{reinhold_error-corrected_2020}, but do not induce additional decay in the storage mode \cite{sun_tracking_2014}. Measurements of $\hat{\mathcal{O}}$ are thus QND, enabling the measurement of multiple generalized parity operators that form the basis of our number-resolving measurement.

\begin{figure*}
	\includegraphics[scale=1]{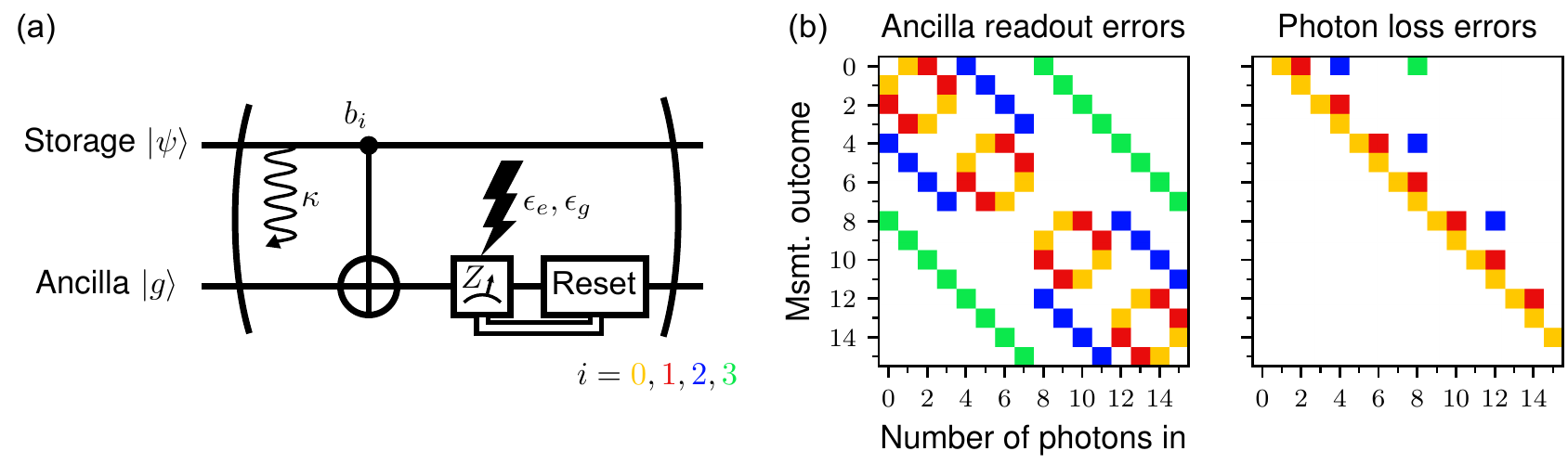}
	\caption{\label{fig2} Measurement circuit diagram and error syndromes. (a) The photon number of a state $\ket{\psi}=\sum_n c_n\ket{n}$ in the storage cavity with decay rate $\kappa$ is measured by sequentially interrogating the least-significant to most-significant bit in the binary decomposition of the photon number ($i = 0, 1, 2, 3$). Optimal control pulses excite the ancilla conditioned on the value of $b_i$ in the storage cavity, followed by a dynamic reset of the ancilla to ensure that it starts in its ground state for the subsequent bit measurement \cite{elder_high-fidelity_2020} (see Appendix \ref{dynamic ancilla reset statistics appendix} for reset statistics). The combined map-measurement process has associated error rates $\epsilon_{e(g)}$ for mis-assigning the ancilla to be in the $\ket{g}(\ket{e})$ state when it should be $\ket{e}(\ket{g})$. (b) Single ancilla (left panel) and decay (right panel) errors associated with the different bits (0 – yellow, 1 – red, 2 – blue, and 3 – green) produce different, but partially overlapping, error syndromes. Note that the syndromes for the decay errors depend on the order in which the bits are measured, and that the entries are qualitative labels of the measurement errors occurred and are not indicative of error magnitudes. }
\end{figure*}
\subsection{Generalized parity measurements}
To implement the number-resolving measurement, we synthesize gates that map the binary-valued generalized parity operators onto the ancilla qubit. Each gate enables the measurement of one bit, so to resolve the first $N$ states, we require $B=\log_2 N$ gates. The generalized parity operators
\begin{equation}
\label{parity_definition}
\left(\hat{\mathcal{P}}^k\right)_{ij}=
\begin{cases}
0 &\text{for}\, i\neq j\\
1-2\left(\left\lfloor\frac{i}{2^k}\right\rfloor(\text{mod}\; 2)  \right) &\text{for}\, i= j
\end{cases}
\end{equation}
with eigenvalues $\lambda=\pm1$ halve the Hilbert space into two eigenspaces $E_{+1}=\{n|b_k(n)=0\}$ and $E_{-1}=\{n|b_k(n)=1\}$, where $b_k(n)$ is the $k^\text{th}$ bit in the binary decomposition of $n$. The $\hat{\mathcal{P}}^k$ are diagonal in the energy eigenbasis and thus satisfy both conditions for QND measurement. To perform this measurement, we use numerical optimal control techniques (see Appendix \ref{OCT appendix} for more details) to synthesize a CNOT-like unitary operation 
\begin{eqnarray}
\label{general_cnot}
\text{C}\hat{\mathcal{P}}^k=&&\sum\limits_{b_k(n)=0}e^{i\varphi_n}\ketbra{n}{n}\otimes\ketbra{g}{g}\nonumber\\&&+\sum\limits_{b_k(n)=1}e^{i\varphi_n}\ketbra{n}{n}\otimes\ketbra{e}{g}\nonumber\\&&+\sum\limits_{b_k(n)=1}e^{i\varphi_n'}\ketbra{n}{n}\otimes\ketbra{g}{e}.
\end{eqnarray}
The gate imparts phases $\varphi_n,\varphi_n'$, but these do not affect the final measurement outcome because the composite measurement projects into a single Fock state, rendering the induced phase irrelevant. We can set these phases ourselves, but here allow the optimizer flexibility to choose the phases. Applying this gate to an ancilla prepared in $\ket{g}$ coupled to a storage mode with state $\ket{\phi}=\sum_n c_n\ket{n}$ entangles the odd and even subspaces with the state of the ancilla,
\begin{eqnarray}
\label{state transfer}
\text{C}\hat{\mathcal{P}}^k(\ket{\phi}\otimes\ket{g})=&&\sum_{b_k(n)=0}c_ne^{i\varphi_n}\ket{n}\otimes\ket{g}\nonumber\\&&+\sum_{b_k(n)=1}c_ne^{i\varphi_n}\ket{n}\otimes\ket{e}.
\end{eqnarray}
QND measurements of the ancilla then project the storage state into either $E_{+1}$ or $E_{-1}$ depending on the outcome, realizing the measurement operators
\begin{equation}
\hat{B}^{(k)}_{0(1)}=\sum\limits_{b_k(n)=0(1)}e^{i\varphi_n}\ketbra{n}{n}.
\end{equation}
The act of measuring $B$ parity operators $\hat{\mathcal{P}}^{B-1},\ldots,\hat{\mathcal{P}}^{1},\hat{\mathcal{P}}^{0}$ with outcomes $b_{B-1},\ldots,b_{1},b_{0}$ in the subspace including only the first $N$ states is equivalent to applying the measurement operator
\begin{equation}
\label{bitwise_meas_op}
\hat{M}_n=\hat{B}^{(B-1)}_{b_{B-1}}\ldots\hat{B}^{(1)}_{b_1}\hat{B}^{(0)}_{b_0}=\ketbra{n}{n}
\end{equation}
and renormalizing. This operator projects the system into the Fock state $\ket{n}=\ket{b_{B-1}(n)\ldots b_1(n)b_0(n)}$ completing the ``bitwise" measurement. This approach requires the minimal number of binary measurements of any scheme assuming no prior knowledge of the state. This protocol implements a detector resolving up to $2^B$ photons, where the number of generalized parity (bit) measurements we make is fully programmable.

\subsection{Experimental implementation}
\label{implementation}
To resolve the storage mode photon number in a single shot, we sequentially measure the generalized parity operators, whose outcomes form the binary decomposition of the photon number. For each bit, we apply $\text{C}\hat{\mathcal{P}}^k$, read out the ancilla, and dynamically reset it using the same method as in Ref.~\cite{elder_high-fidelity_2020}. More details of the reset protocol, including statistics on the number of attempts required, are in Appendix \ref{dynamic ancilla reset statistics appendix}. This process is concatenated to measure all four bits, allowing us to resolve up to 15 photons, as shown in Fig.~\ref{fig2} (a).

Another approach performs a series of parity-like measurements, feeding forward the result of each bit measurement to determine the correct qubit rotation angle for the next measurement \cite{dassonneville_number-resolved_2020}. However, any errors that occur during a single measurement are fed forward and corrupt any subsequent measurements, resulting in correlated errors. Our method, which independently measures each bit, is more suitable for error mitigation with post-processing techniques.

Each component of our measurement chain is susceptible to errors. The storage mode suffers from stochastic photon loss at a rate $\kappa$. The ancilla is vulnerable to dephasing and excitation decay during the mapping pulse $\text{C}\hat{\mathcal{P}}_N$ as well as decay during the readout procedure. The probability of reading out the ancilla in the ground (excited) state when the readout should have yielded $e(g)$ is $\epsilon_{e(g)}$. The long lifetimes of the storage mode $T_1^s\approx1\,\text{ms}$ and ancilla $T_1^\sigma\approx25\,\mu\text{s}$ relative to the $2.9\,\mu\text{s}$ duration of each measurement ensure that these errors are small. The dynamic reset protocol rarely leaves the ancilla in the excited state, and is not a significant source of error.

Counter-intuitively, a single error in the entire sequence can produce a result that differs from the correct value by up to eight photons. This happens because a single error can flip multiple bits in the photon number's binary decomposition. For example, the loss of a single photon from $\ket{8}=\ket{1000}$ after the measurement of $b_0=0,b_1=0,b_2=0$ results in $b_3(7)=0$. Thus, the effect of such an error is to mistakenly read out $\ket{8}=\ket{1000}$ as $\ket{0}=\ket{0000}$. This error is illustrated by the green square in the right panel of Fig.~\ref{fig2} (b), which also shows the measurement outcomes when a single ancilla readout or storage decay error occurs.

\section{Measurement Error Mitigation}
\label{sec_3}
Can we use our knowledge of the error mechanisms described above to improve the fidelity of our measurements? To answer this question, we use an approach known as error mitigation which relies on data post-processing to improve the quality of an ensemble of calculations or measurements. Error mitigation has been proposed \cite{temme_error_2017, li_efficient_2017} and realized \cite{kandala_error_2019} for use with variational algorithms. Here, our error mitigation efforts focus on an ensemble of measurements which sample from the population distribution of some quantum state.

In this section, we introduce an error model based on the error syndromes described in Sec.~\ref{implementation}. This model is a function of several error rates, which must be calibrated. We discuss our calibration technique and compare the model to experimental results. Finally, we introduce the deconvolution techniques that use the calibrated model to mitigate the errors in the measurement and show an improvement on two sets of states.

\begin{figure*}[t]
	\centering
	\includegraphics[scale=1]{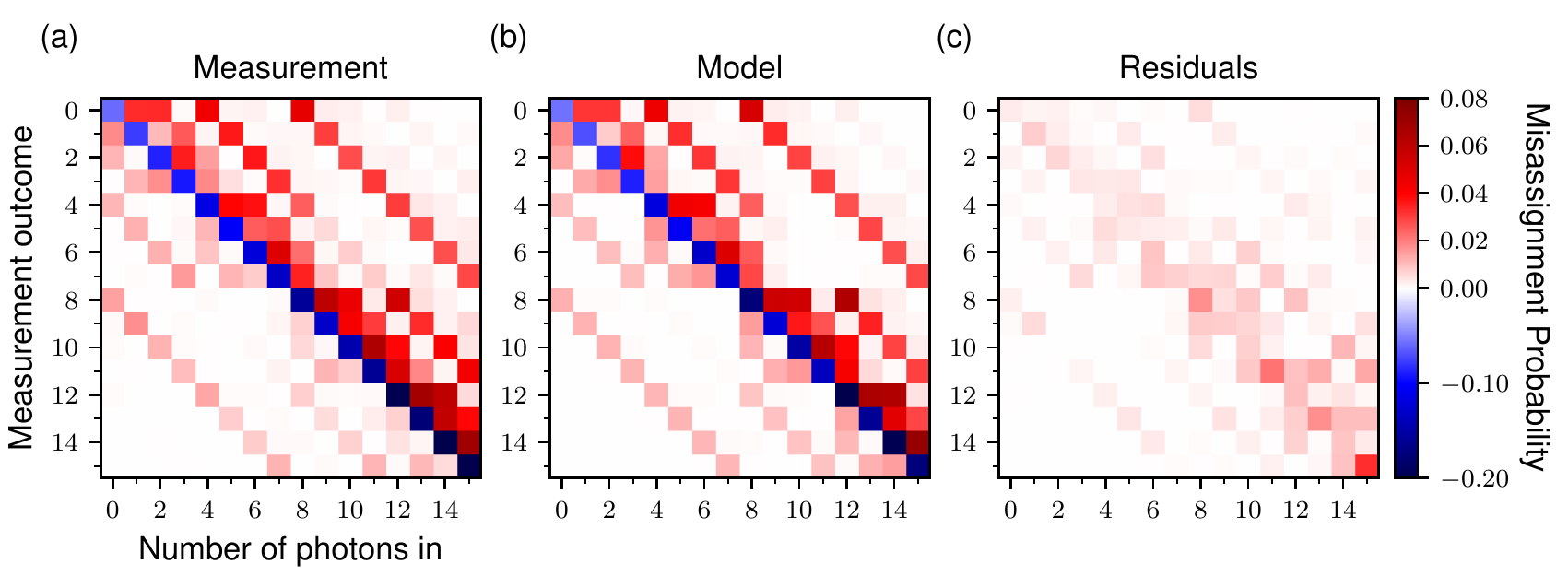}
	\caption{\label{fig3} Measured error syndromes and hidden Markov model prediction. (a) Misassignment probabilities of prepared Fock states measured bitwise. (b) Misassignment probabilities predicted by the HMM. The HMM models the error syndromes depicted in Fig.~2 using the known storage mode loss rate $\kappa$ and the calibrated error rates $\epsilon^{(k)}_g,\,  \epsilon^{(k)}_e$. The HMM correctly predicts the structure of the errors (b) and agrees with the data down to the percent level (c). The negative values along the diagonal indicate the correct assignment probability deficit with each column summing to zero. The rows are the measured and modeled $\{F_i-\mathbf{1}_i\}$, respectively, where $\mathbf{1}_i$ is the matrix that is zero except for a one in its $i^\text{th}$ diagonal entry. The residuals may arise from calibration errors or other error syndromes that are not modeled by the HMM. The residuals provide a typical bound of less than $1\%$, and no more than $4\%$, on the error syndromes not included in the HMM. }
\end{figure*}
\subsection{Error model}
Our goal is to characterize the errors so that we can model measurement outcomes. This is equivalent to finding the elements $\{F_i\}_{0\leq i\leq N_{\text{max}}}$ of the positive operator-valued measure (POVM) that describes our measurement \cite{nielsen_quantum_2011, maciejewski_mitigation_2020}. For instance, the bitwise measurement is composed of $B$ individual bit measurements, each with a corresponding POVM $\{E^{(k)}_{b_k=0},E^{(k)}_{b_k=1}\}$ where $E^{(k)}_{b_k}=\left(\hat{B}_{b_k}^
{(k)}\right)^\dagger \hat{B}_{b_k}^{(k)}$. Each ideal $F_i$ is then a product of POVM elements constituting the bitwise measurement:
\begin{equation}
F_{i=b_{B-1}\ldots b_0}=E^{(B-1)}_{b_{B-1}}\ldots E^{(0)}_{b_0}.
\end{equation}
Note that the $F_i$ and $E^{(k)}_{b_k}$ are diagonal because the bit measurement operators $\hat{B}_{b_k}^{(k)}$ are QND. Ideally, $F_i=\ketbra{i}{i}$, but our implementations of the $E^{(k)}_{b_k}$ have errors, so $\left(F_i\right)_{j,j}$ is the probability that the measurement detects $i$ photons when there are $j$ photons in the mode $P(\text{outcome}=i|\ket{j})$.

Our system can be described with a hidden Markov model (HMM) that has been used for qubit readout \cite{dreau_single-shot_2013, gammelmark_past_2013, ng_optimal_2014, wolk_state_2015, martinez2020improving} and to improve the readout of qubits encoded in oscillators \cite{hann_robust_2018,elder_high-fidelity_2020}. A HMM parametrizes the errors in the storage mode, known as transitions, and the fidelity of the measurements of each bit, known as emissions. The probability of a transition in the storage mode during a time interval $t$ from $\ket{i}$ to $\ket{j}$ is
\begin{equation}
T_{i,j}\left(\kappa t\right)=\binom{i}{j}\left(e^{\kappa t} - 1\right)^{i-j}e^{-i\kappa t},
\end{equation}
where $\kappa$ is the decay rate of the mode \cite{hann_robust_2018}. The ancilla can decay during the controlled rotation $\text{C}\hat{\mathcal{P}}^k$ and the readout procedure, leading to an incorrect bit measurement. Since the bit measurements are QND, only the diagonal entries $E_{b_k,i}^{(k)}$ are non-zero, even in the presence of errors. We modify the entries of these ideal POVM elements
\begin{equation}
E_{b_k,i}^{(k)}=
\begin{cases}
\left\lfloor\frac{i}{2^k}\right\rfloor \;\text{is even}\;
&\begin{cases}
1-\epsilon_g^{(k)}\;&\text{ for }\; b_k=0 \\
\epsilon_g^{(k)}\;&\text{ for }\; b_k=1
\end{cases}\\
\left\lfloor\frac{i}{2^k}\right\rfloor \;\text{is odd}\;
&\begin{cases}
\epsilon_e^{(k)}\;&\text{ for }\; b_k=0 \\
1-\epsilon_e^{(k)}\;&\text{ for }\; b_k=1
\end{cases}
\end{cases}
\end{equation}
to introduce error rates $\epsilon_{g(e)}$ describing errors in the $\text{C}\hat{\mathcal{P}}^k$ mapping or ancilla readout.

We now integrate both error mechanisms into a single model describing the errors in the measurement. The bitwise measurement algorithm described in Sec.~\ref{implementation} alternates between entangling pulses and ancilla reset. The cavity is subject to spontaneous decay for the duration of the algorithm, punctuated by projective measurements of the entangled ancilla. This duration, however, is not deterministic due to the dynamic ancilla reset. Our HMM models the errors in the measurement by constructing an alternating chain of transition and emission (measurement) events. To calculate the $\{F_i\}$, we sum over all transition paths $s_k$ between storage state $\ket{j}$ and measurement outcome $i$. Weighting these paths by the probability that the bit measurements yield the binary decomposition of $i=b_3b_2b_1b_0$ gives
\begin{widetext}
\begin{equation}
\label{C_def}
\left(F_i\right)_{j,j}=\sum\limits_{s_1\ldots s_{B}}T_{j, s_1}\left(\kappa t^{(0)}\right)E^{(0)}_{b_0,s_1}T_{s_1, s_2}\left(\kappa t^{(1)}+b_0\kappa t'\right)E^{(1)}_{b_1,s_2}\ldots T_{s_{B-1}, s_B}\left(\kappa t^{(B-2)}+b_{B-1}\kappa t'\right)E^{(B-1)}_{b_{B-1},s_B}.
\end{equation}
\end{widetext}
Each transition matrix $T$ and POVM element $E^{(k)}_i$ are parametrized by the respective error rates $\kappa t^{(k)}$ and $\epsilon_{g}^{(k)}, \epsilon_{e}^{(k)}$, which are not necessarily the same for each bit. Note that the time between each bit measurement depends on each result, as a measurement of $b_k=1$ requires an ancilla reset pulse followed by readout verification. We discuss how the dynamic ancilla reset statistics inform our choice of $\kappa t'$ in Appendix \ref{dynamic ancilla reset statistics appendix}. Eq. \ref{C_def} provides the desired parametrized POVM elements, which, once calibrated, we  use to mitigate errors in our measurement.

\subsection{Model calibration and results}
\label{model calibration}
Now that we have a model for the errors, we must determine the values of the error parameters $\kappa t^{(k)}$ and $\epsilon_{g}^{(k)}, \epsilon_{g}^{(k)}$. The duration of each bit measurement is recorded shot-by-shot; the microwave pulses have fixed durations and the dynamic ancilla reset records the number of required attempts. Since measurements of the cavity's decay rate $\kappa$ are straightforward, we only need to determine the $\epsilon_{g}^{(k)}, \epsilon_{g}^{(k)}$.

We calibrate the bit measurement errors by measuring a single bit on a basis of states. This characterizes the $\{E^{(k)}_i\}$ from which we extract the $\epsilon_{g}^{(k)}, \epsilon_{g}^{(k)}$. This procedure relies on the ability to prepare a basis of states with high fidelity, so that preparation errors do not pollute the measurement errors. We prepare Fock states by repeatedly converting two excitations in the ancilla into a single photon as detailed in \cite{gasparinetti_measurement_2016,elder_high-fidelity_2020}. We check the photon number with a series of selective pulses. If any fail to flip the ancilla, the storage mode is cooled to vacuum and the preparation protocol tries again. This method prepares high-fidelity Fock states, but does allow photon decay during the final check (see the Supplementary Material for \cite{elder_high-fidelity_2020}). Appendix \ref{choice of calibration states appendix} contains additional details about the calibration process, including how we account for these preparation errors.

Once we have calibrated the error rates in the model (tabulated in Appendix \ref{model error rates appendix}), we use Eq. \ref{C_def} to calculate the $\{F_{i}\}$ and thus reconstruct the detector POVM. We compare the modeled POVM to the measured POVM in Fig.~\ref{fig3} by preparing the Fock states $\{\ket{j}\}_{0\leq j\leq 15}$ and measuring all four bits. The residuals are typically $~1\%$, and never more than $4\%$, showing good agreement with the model and bounding unmodeled errors. We could in principle use the measured detector POVM to perform error mitigation, but we would still need to model transitions during the Fock state preparation and would not be able to confirm our understanding of the error mechanisms.

The raw measurement fidelity of each state is at least $80\%$ before error mitigation even for the most challenging input state $n=15$. The measurement extracts on average between 3.11 and 3.17 of the four possible bits of information (the range exists due to state preparation details; see Appendix \ref{efficiency calculation} for further discussion). This fraction is greater $3/4$, suggesting that the bitwise measurement is single-shot.
\begin{figure*}
	\centering
	\includegraphics[scale=1]{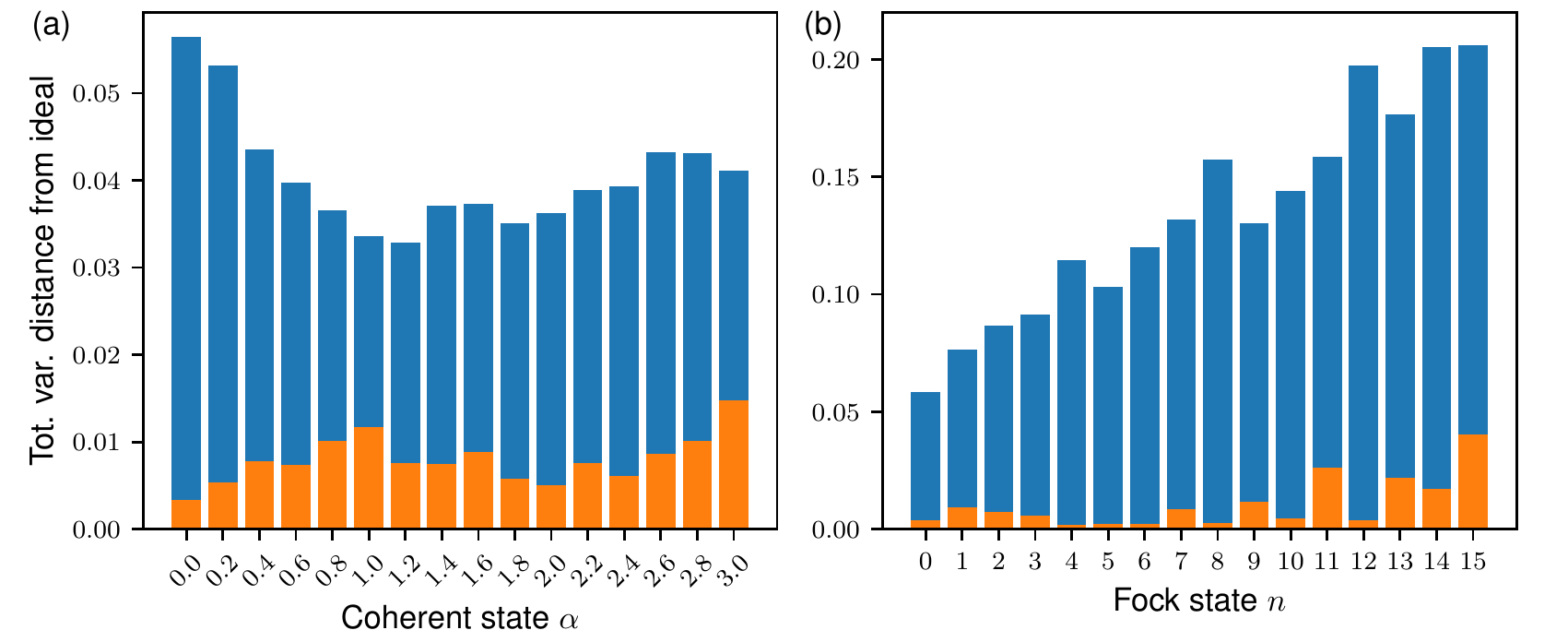}
	\caption{\label{fig4} Error mitigation results for coherent (left panel) and Fock states (right panel). The $\alpha$ were chosen such that the Hilbert space can be truncated to the first 16 Fock states, with error less than 2.2\%. For each prepared state $\ket{\psi}$, the error in the measurements $D_{TV}\left(\textbf{P}_{\ket{\psi}}^{\text{meas}}, \textbf{P}_{\ket{\psi}}^{\text{ideal}}\right)$ (blue) is compared to the error in the mitigated measurements $D_{TV}\left(\textbf{P}_{\ket{\psi}}^{\text{mit}}, \textbf{P}_{\ket{\psi}}^{\text{ideal}}\right)$ (orange). The large difference in overall error rates between Fock and coherent states results from the error mechanism to which each is most susceptible. Fock states transition to orthogonal states under photon loss, while coherent states are mainly affected by bit measurement errors. The single photon loss probability $\kappa t^{(k)}\approx0.03$ is similar to the single bit error rates $\epsilon_{e}^{(k)}\approx.03,\epsilon_{g}^{(k)}\approx.01$.  } 
\end{figure*}
\subsection{Error deconvolution}
Now that we have modeled the entire collection of $\{F_i\}$, we form a matrix $C_{ij}=\left( F_i \right)_{j,j}$, where $C-I$ is shown in Fig.~\ref{fig3}(b), of conditional probabilities referred to as the ``confusion matrix." This allows us to relate the measured populations $\textbf{P}^{\text{meas}}$ to the state's ideal distribution $\textbf{P}^{\text{ideal}}$ with a simple relation \cite{maciejewski_mitigation_2020}
\begin{equation}
\label{confusion_matrix}
\textbf{P}^{\text{meas}}=C\textbf{P}^{\text{ideal}}.
\end{equation}
Once we have $C$, we invert it to solve the above equation for the mitigated probability vectors $\textbf{P}^{\text{mit}}=C^{-1}\textbf{P}^{\text{meas}}$ (Appendix \ref{C_invertibility} contains a discussion of the invertibility of $C$). These vectors are properly normalized, but may have negative entries. We correct the mitigation results by finding the closest vector $\textbf{P}^{\text{mit}}$ with respect to the Euclidean norm that has non-negative entries \cite{maciejewski_mitigation_2020}.

The simple form of Eq. \ref{confusion_matrix} offers the tantalizing possibility that we could avoid error modeling altogether by preparing a high-fidelity set of states, and directly inverting $\textbf{P}^{\text{ideal}}$ to find $C$. Coherent states seem to be an ideal candidate due to the high fidelity with which we can prepare them, but suffer from ill-conditioned ideal populations $\textbf{P}^{\text{ideal}}$, which make the determination of $C$ highly sensitive to errors in $\textbf{P}^{\text{meas}}$ \cite{Lundeen2009}. Furthermore, error modeling of the Fock states is required due to preparation error discussed in the previous section.

To quantify the quality of our measurement, we use the total variation distance metric to compare probability distributions $\textbf{A}$ and $\textbf{B}$,
\begin{equation}
\label{L1}
D_{TV}(\textbf{A},\textbf{B})=\frac{1}{2}\sum\limits_{k} \left| \textbf{A}_k-\textbf{B}_k\right|.
\end{equation}
The factor of $1/2$ accounts for the double-counting of probability differences.

To check the performance of the error mitigation and verify the error model, we apply it to the same Fock states we prepared in the calibration process, and to a set of coherent states with $0\leq\alpha\leq3$. The results are shown in Fig.~\ref{fig4}. In both cases, the error mitigation includes photon losses occurring during state preparation (recall the selective pulse check discussed in Sec.~\ref{model calibration}). This is reasonable because the operation that directly precedes the bitwise measurement is known in advance as part of the pulse sequence.

The large difference in the pre-mitigated errors in Fig.~\ref{fig4} between the Fock and coherent states is due to the nature of these states. For the Fock states, a single lost photon transitions $\ket{n}$ to an orthogonal state $\ket{m}$, resulting in the maximum possible error. Furthermore, the transition rate increases with photon number, resulting in large pre-mitigation errors. On the other hand, coherent states are more robust to photon loss and do not transition to an orthogonal state, leaving bit measurement errors as the dominant error mechanism.

The error mitigation successfully reduces errors in all tested states to less than $5\%$, providing the most dramatic improvement to the Fock states, whose average error is reduced from $13.5\%$ to $1.1\%$. The success of the error mitigation shows that we understand the error syndromes in our measurement and can effectively quantify them. This allows us to use the model to find the error rates required to meet a desired measurement fidelity, which may vary from application to application.

\subsection{Scalability of error mitigation}

In the previous sections, we demonstrated an efficient measurement of the number of photons in a mode, and an effective method to mitigate the errors in the measurement. Our protocol can be performed simultaneously on a system with multiple storage modes, each dispersively coupled to their own ancilla qubit, to measure the joint photon number of the system. Imagine a system with $M$ storage modes with uncorrelated measurement errors between the modes. Here, the entries of $\mathbf{P}^{\text{meas}}$ are elements of the joint probability distribution $P(n_{1},n_{2},\ldots,n_{M})$. If we truncate the Hilbert space of each mode to some maximal number of photons $N_{\text{max}}$, then the length of $\mathbf{P}^{\text{meas}}$ is $N_{\text{max}}^M$. This exponentially large space suggests that our approach to error mitigation is not scalable to multiple modes. 

However, it turns out that calculating any particular element of $\mathbf{P}^{\text{mit}}$ from $\mathbf{P}^{\text{meas}}$ is efficient, as we show below. Calculating particular elements of $\mathbf{P}^{\text{mit}}$ may be sufficient for problems in which the goal is to study features of an output distribution that can be altered by measurement errors. For instance, we may be interested in correcting the relative peak intensities of Franck-Condon factors \cite{wang_efficient_2020, clements_approximating_2018}, which only requires processing the measured peaks, not the entire Hilbert space. The resulting spectrum will have significant peaks at the same output photon numbers, with intensities adjusted to account
for measurement errors, increasing the simulator's accuracy.
This way, the simulator still does all the hard work
of identifying the significant peaks; we only need to post-process the heights to mitigate measurement errors.

There are two properties that make calculating any element of $\mathbf{P}^{\text{mit}}$ efficient. The first is that the multi-mode confusion matrix is the Kronecker product of single-mode confusion matrices, meaning that it is efficient to calculate any element in the inverse of the multi-mode confusion matrix. In general,
though, $C^{-1}$ contains an exponential number of nonzero matrix
elements so writing down the entire matrix is not feasible. The second
property is that the vector $\mathbf{P}^{\text{meas}}$ is necessarily sparse;
assuming we do not perform an exponentially large number of measurements,
most configurations $(n_{1},n_{2},\ldots,n_{M})$ are never actually
measured. Together, these two properties imply that any entry of $\mathbf{P}^{\text{mit}}$
can be efficiently computed, 
\begin{equation}
\label{peak by peak}
\mathbf{P}_{i}^{\text{mit}}=\sum_{j\in\mathcal{S}}\left(C_{1}^{-1}\otimes C_{2}^{-1}\otimes\ldots\otimes C_{M}^{-1}\right)_{ij}\mathbf{P}^{\text{meas}}_{j}
\end{equation}
where $\mathcal{S}$ is the set of nonzero entries in $\mathbf{P}^{\text{meas}}$,
which is only polynomially large. We cannot use this fact to efficiently
calculate the entire distribution $\mathbf{P}^{\text{mit}}$, however, because
$\mathbf{P}^{\text{mit}}$ can contain exponentially many nonzero entries. In the case where simply adjusting peak intensities is insufficient, we have developed a method for expanding $C^{-1}$ that mitigates errors up to a chosen order. This method is detailed in Appendix \ref{confusion matrix expansion appendix}.

\section{Conclusion and outlook}
Our bitwise measurement protocol utilizes the long lifetimes of 3D microwave cavities and the interactions generated by a Josephson junction to implement a high-fidelity, single-shot, microwave photon number-resolving detector. We use error mitigation techniques to improve the measurement fidelity of the Fock states by nearly an order of magnitude. Larger resolvable photon numbers are reachable by synthesizing additional $\text{C}\hat{\mathcal{P}}^k$ pulses to measure more bits.

In addition to enabling error mitigation, our error model provides insight into the error budget of our PNR detector. For example, we estimate that with unit fidelity $\text{C}\hat{\mathcal{P}}^k$ pulses and ancilla readout ($\epsilon_{g}^{(k)}=\epsilon_{e}^{(k)}=0$) we can extract $3.72$ bits of information in a single shot in our system with the current $\kappa t^{(k)}$. This suggests that further optimization of the control pulses and ancilla readout can increase the information extracted per shot by up to $0.6$ bits. Any additional improvements to the single-shot fidelity require decreasing $\kappa t^{(k)}$ by decreasing the pulse lengths or increasing cavity lifetimes. Adaptive check methods may also be able to improve single-shot fidelity at the expense of adding complexity to the error model.

Even with limited fidelity, optical PNR detectors have motivated the development of multiphoton quantum information processing protocols. Bringing high-fidelity PNR detectors to the microwave regime further bolsters the ability of cQED to perform photonic protocols, such as vibronic spectra simulations \cite{wang_efficient_2020}. Additionally, this capability will motivate the development of algorithms that take advantage of the large bosonic Hilbert space available in cQED systems.

\begin{acknowledgments}
	We acknowledge Reinier Heeres for originally proposing to measure photon number with its binary decomposition. We thank Serge Rosenblum and Philip Reinhold for experimental support and N. Frattini for providing the Josephson
	parametric converter. This research was supported by the
	U.S. Army Research Office through grant W911NF-18-1-0212. Fabrication
	facilities use was supported by the Yale Institute for Nanoscience and
	Quantum Engineering (YINQE) and the Yale SEAS clean room. L.F. and
	R.J.S. are co-founders of, and equity shareholders in, Quantum Circuits,
	Inc. C. T. H. acknowledges support from the NSF GRFP (DGE1752134). L.J. acknowledges support from the Packard Foundation (2013-39273).
\end{acknowledgments}

\appendix

\section{Dynamic ancilla reset statistics}
\label{dynamic ancilla reset statistics appendix}
\begin{figure}
	\centering
	\includegraphics[scale=1]{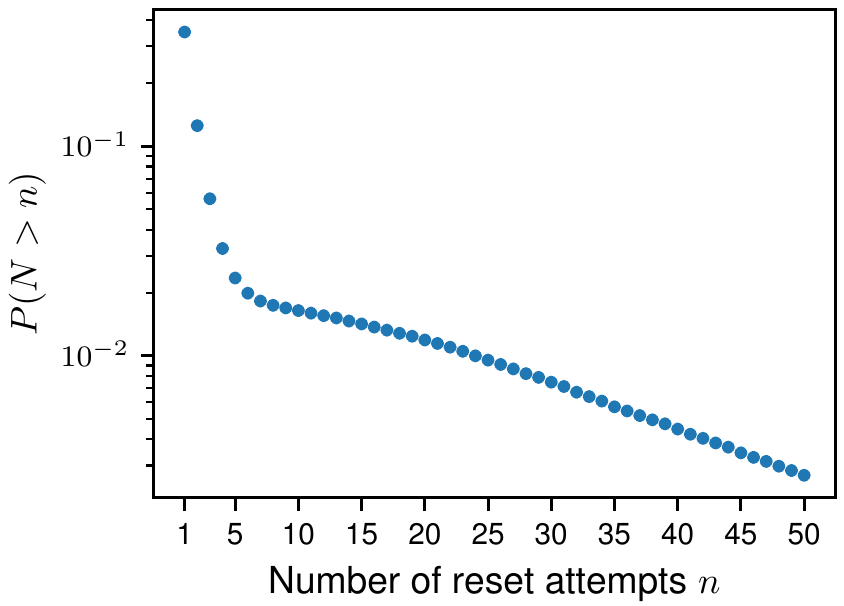}
	\caption{\label{supp_fig1} Ancilla dynamic reset probabilities from the Fock state data in Figs. \ref{fig3} and \ref{fig4}(b). After each bit measurement, we readout and, if necessary, dynamically reset the ancilla. This protocol requires more than one attempt with probability $P(N>1)=0.351$ and at most 50 with probability $1-P(N>50)=0.997$. } 
\end{figure}

As discussed in the supplemental material in \cite{elder_high-fidelity_2020}, our ancilla qubit reset protocol does not always succeed in a single attempt. This increases the duration between bits on a shot-by-shot basis. This duration is proportional to the number of reset attempts, whose probabilities shown in Fig.~\ref{supp_fig1} have a long tail, with $35.1\%$ of resets requiring more than one attempt. The average number of reset attempts required across all Fock states prepared in Figs. \ref{fig3} and \ref{fig4}(b) is $\bar{n}_{\text{reset}}=2.05$. Each reset attempt has a duration of $2.244\,\mu\text{s}$ so we use the average reset time $\kappa t'=2.244\,\mu\text{s}\cdot\kappa\bar{n}_{\text{reset}}=0.0046$ in the model.

It is tempting to remove runs via postselection that take more than a chosen threshold of reset attempts, but this introduces bias into the measurement results as the number of required reset attempts increases with the number of photons in the storage mode. Postselecting on a low threshold such as five attempts (used in \cite{elder_high-fidelity_2020}) removes $5\%$ more measurements of $\ket{15}$ than $\ket{1}$. Setting the postselection threshold at 50 attempts lowers this disparity to $0.6\%$, but we still choose to avoid postselection to produce an unbiased measurement. A more granular approach to ancilla reset using the state-dependent reset probability distributions can be included in the HMM if one separately calibrates this dependence.

\section{Choice of calibration states}
\label{choice of calibration states appendix}
Consider a set of calibration states $\{\ket{\psi_j}\}$. After preparing each $\ket{\psi_j}$, we measure the $k^{\text{th}}$ bit to find the probability of measuring $b_k=0,1$ for each basis state $\textbf{P}^{\text{cal}}_{b_k, j}=P(b_k=0,1|\ket{\psi_j})=\braket{\psi_j|E^{(k)}_{b_k}|\psi_j}$. Our preparation protocol may suffer from transitions caused by ancilla readout and storage mode decay that we can model in the calibration process
\begin{equation}
\label{E_cal_sum}
\textbf{P}^{\text{cal}}_{b_k, j}=\sum\limits_{m,n}T_{m,n}(\kappa t^{(0)})E^{(k)}_{b_k,n}O_{j,m},
\end{equation}
where $O_{j,m}=|\braket{\psi_j|m}|^2$ changes from the calibration basis $\{\ket{\psi_j}\}$ to the Fock basis. The argument of $T$, $\kappa t^{(0)}$, includes the duration of the final check of the preparation protocol and the $C\hat{\mathcal{P}}^0$ pulse. It can be shown that the effective transition duration for this final check is approximately half of the length of the selective pulse. The value of $\kappa t^{(0)}$ is given in Table \ref{measured_error_parameters}. Solving for the diagonal of $E^{(k)}_{b_k}$ we find
\begin{equation}
\label{E_cal}
E^{(k)}_{b_k,i}=\left(T^{-1}(\kappa t^{(0)})O^{-1}\textbf{P}^{\text{cal}}\right)_{i,i}
\end{equation}
from which we average and extract $\epsilon_{g}^{(k)}, \epsilon_{g}^{(k)}$.

Our task is then to pick a basis of calibration states $\ket{\psi_j}$. In our systems, we can prepare coherent states $\ket{\alpha}$ with the highest fidelity due to the speed of preparation. If we choose a calibration basis $\{\ket{\alpha_j}\}$ such that there is very little probability finding $n$ photons with $n>N_{\text{max}}$, we would expect these states to be an excellent candidate for a calibration basis. However, the presence of $O^{-1}$ in Eq. \ref{E_cal} complicates matters. To satisfy $n\leq N_{\text{max}}$, the range of $\alpha_j$ is necessarily restricted. In this regime, the coherent states are only barely linearly independent and have significant co-overlap $\braket{\alpha_j|\alpha_k}$. This results in $O$ being an ill-conditioned matrix, meaning that $O^{-1}$ magnifies small errors in $\textbf{P}^{\text{cal}}$ \cite{maciejewski_mitigation_2020}. We avoid the ill-conditioning problem by using the Fock states $\{\ket{j}\}_{0\leq j\leq N_{\text{max}}}$ as our calibration basis resulting in $O=1$.
\section{Model error rates}
\label{model error rates appendix}
Table \ref{measured_error_parameters} contains error parameters calibrated in Sec.~\ref{model calibration} and used in the HMM.
\begin{table}[h]
	\caption{\label{measured_error_parameters}%
		HMM error parameters. The error $\sigma$ of $\epsilon_{g}^{(k)},\epsilon_{e}^{(k)}$ is calculated by propagating error in Eq. \ref{E_cal}. $1/\kappa\approx1\text{ms}$ is the storage cavity decay rate.
	}
	\begin{ruledtabular}
		\begin{tabular}{cccc}
			\textrm{Bit $k$}&
			\textrm{$\kappa t^{(k)}\,(+\kappa t')\footnotemark[2]$}&
			\textrm{$\epsilon_{g}^{(k)}\pm\sigma$}&
			\textrm{$\epsilon_{e}^{(k)}\pm\sigma$}\\
			\colrule
			0 & 0.0040, 0.0032\footnotemark[1]  & $0.019\pm0.002$ & $0.029\pm0.002$\\
			1 & $0.0034\,(+0.0046)\footnotemark[2]$ & $0.014\pm0.001$ & $0.026\pm0.001$\\
			2 & $0.0034\,(+0.0046)\footnotemark[2]$ & $0.011\pm0.001$ & $0.035\pm0.002$\\
			3 & $0.0034\,(+0.0046)\footnotemark[2]$ & $0.013\pm0.001$ & $0.033\pm0.002$\\
		\end{tabular}
	\footnotetext[1]{Fock and coherent states, respectively, which have different preparation methods}
	\footnotetext[2]{The parenthetical additional time is added when the ancilla needs to be reset and is equal to $\kappa t'$, as discussed in Appendix \ref{dynamic ancilla reset statistics appendix}.}
	\end{ruledtabular}
\end{table}

\begin{figure*}
	\centering
	\includegraphics[scale=1]{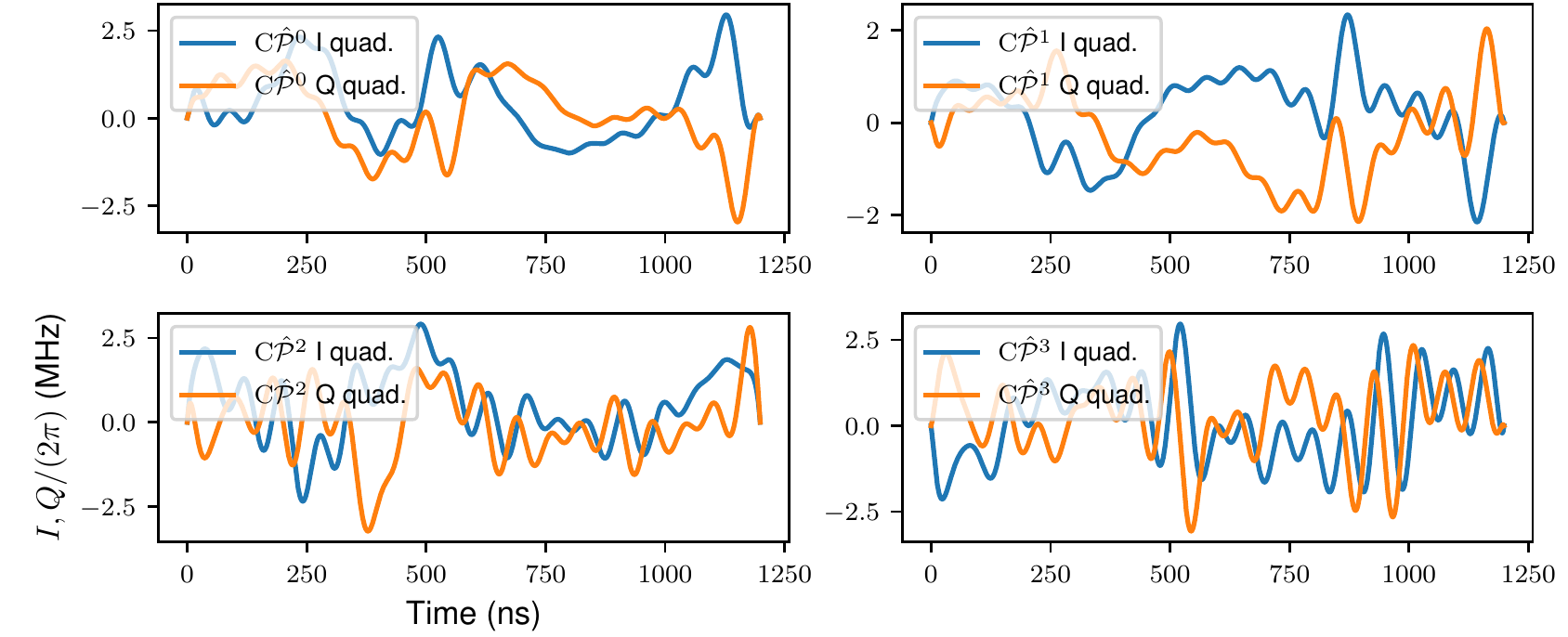}
	\caption{\label{supp_fig2} $\text{C}\hat{\mathcal{P}}^k$ pulses generated by optimal control. Each pulse has a fixed duration of 1200 ns. The amplitude is multiplied by a calibrated constant before being played by FPGA-based signal synthesizers. The signal is upconverted via an IQ mixer and passed through a series of amplifiers and attenuators before reaching the ancilla qubit. } 
\end{figure*}

\section{$\text{C}\hat{\mathcal{P}}^k$ pulse construction}
\label{OCT appendix}
Optimal control techniques are a powerful tool to engineer quantum gates and initialize non-classical quantum states. We use this tool here to create the $\text{C}\hat{\mathcal{P}}^k$ entangling gates in a manner very similar to that described for a cavity-transmon system in Ref.~\cite{heeres_implementing_2017}.

In particular, we use Gradient Ascent Pulse Engineering (GRAPE) \cite{khaneja_optimal_control, deFouquieres_2nd_order_GRAPE} to compute both quadratures of an ancilla control pulse that realizes $\text{C}\hat{\mathcal{P}}^k$. The optimizer uses the cavity-qubit Hamiltonian in the rotating frame
\begin{eqnarray}
\label{OCT hamiltonian}
\hat{H}_{\text{GRAPE}}/\hbar=&&-\frac{\alpha}{2} \hat{a}^\dagger\hat{a}^\dagger\hat{a}\hat{a}-\frac{K}{2}\hat{s}^\dagger\hat{s}^\dagger\hat{s}\hat{s}\nonumber\\
&&-\frac{\chi_{s\sigma}}{2}\hat{s}^\dagger\hat{s}\hat{a}^\dagger \hat{a}+\frac{\chi'_{s\sigma}}{2}\hat{s}^\dagger\hat{s}^\dagger\hat{s}\hat{s}\hat{a}^\dagger \hat{a}\nonumber\\
&&+I(t)(\hat{a}+\hat{a}^\dagger)+iQ(t)(\hat{a}-\hat{a}^\dagger)
\end{eqnarray}
to compute the best $\Omega(t)=I(t)+iQ(t)$ that implements $\text{C}\hat{\mathcal{P}}^k$. We convert $\Omega(t)$ into a DAC amplitude with a scaling factor determined by calibrating the Rabi rate of a $\pi$-pulse on the ancilla qubit.  The substitution $\hat{\sigma}_z\rightarrow \hat{a}^\dagger \hat{a}$ allows GRAPE to optimize over multiple transmon levels. We do not include loss in the qubit and cavity in the optimization. The optimizer finds the ideal control pulse $\Omega$ by maximizing the fidelity of the state transfer in Eq. \ref{state transfer}
\begin{equation}
\mathcal{F}(\Omega)= \sum_k \left|\braket{\psi^{(k)}_{\text{target}}|\hat{U}(\Omega)|\psi^{(k)}_{\text{initial}}} \right|^2
\end{equation}
where $\hat{U}(\Omega)$ is the unitary implemented by the pulse $\Omega$. Note that the modulus is taken within the sum, so the optimal pulse produced by the optimization maximizes the incoherent fidelity of the $\text{C}\hat{\mathcal{P}}^k$ gates. Both quadratures of these pulses are shown in Fig.~\ref{supp_fig2} with Hamiltonian parameters listed in Table \ref{OCT_parameters}. Each pulse reaches a fidelity of at least $99.9\%$ in a lossless system.

\begin{table}[h]
	\caption{\label{OCT_parameters}%
		Parameters for the Hamiltonian \ref{OCT hamiltonian} used in the optimal control construction of the $\text{C}\hat{\mathcal{P}}^k$ gates.
	}
	\begin{ruledtabular}
		\begin{tabular}{cccc}
			&\textrm{Parameter}&
			\textrm{Value}& \\
			\colrule
			&$\alpha/(2\pi)$ & 132 MHz&\\
			&$K/(2\pi)$ & 2.59 kHz& \\
			&$\chi_{s\sigma}/(2\pi)$ & 885 kHz& \\
			&$\chi'_{s\sigma}/(2\pi)$ & 3.67 kHz& \\
		\end{tabular}
	\end{ruledtabular}
\end{table}
\section{Calculation of information extracted by the detector}
\label{efficiency calculation}
To be able to call our detector single-shot, it must extract a large fraction of the information it is designed to measure per shot. In particular, of the four bits it ideally measures, we would like to determine how many of these bits are lost to measurement errors. The difference of ideal and wasted bits gives the average number of bits extracted per measurement.

Our task is then to calculate the number of wasted bits in our measurement, which is equivalent to number of extra bits needed to classify the input given a measurement result (or the number of bits remaining to be measured). This quantity is the entropy $S_i$ \cite{shannon_entropy} of the distribution $P(\ket{j}|i)=P(\text{detector input }\ket{j}|\text{measurement result }i)$. We finally average over the results $i$ to obtain $4-\langle S_i\rangle$, the average number of bits extracted in a single shot of the measurement.

We begin by calculating the conditional probability
\begin{eqnarray}
\label{bayes1}
P(\ket{j}|i)=&& \frac{P(\text{result }i|\text{input }\ket{j})P(\ket{j})}{P(i)}.
\end{eqnarray}
Recalling that $P(\text{result }i|\text{input }\ket{j})=C_{ij}$ and assuming a uniform prior $P(\ket{j})=1/N$ we simplify Eq. \ref{bayes1}
\begin{eqnarray}
P(\ket{j}|i)=&& \frac{C_{ij}P(\ket{j})}{\sum_k C_{ik}P(\ket{k})}\nonumber \\
=&&\frac{C_{ij}}{\sum_k C_{ik}}.
\end{eqnarray}
Finally, we write down the average entropy over all measurement outcomes, weighting by $P(\text{outcome } i)$
\begin{eqnarray}
\langle S_i\rangle &&= -\sum\limits_{ij} P(\ket{j}|i)\log_2(P(\ket{j}|i))P(i)\nonumber\\
&&=-\frac{1}{N}\sum\limits_{ij} C_{ij}\log_2\left( \frac{C_{ij}}{\sum_k C_{ik}} \right).
\end{eqnarray}
Using the error parameters listed in Table \ref{measured_error_parameters} to construct $C$, we find $4-\langle S_i\rangle_{C,\text{ Fock}}=3.14$ bits. We can also use the measurement results in Fig.~\ref{fig3} to determine $P(\text{result }i|\text{input }\ket{j})$, yielding $4-\langle S_i\rangle_{\text{Fock}}=3.11$ bits. This number likely provides an upper bound of $\langle S_i\rangle$ as the check protocol for Fock states (described in Sec.~\ref{model calibration}) introduces a lengthy delay between state preparation and measurement.

As mentioned above, the preparation errors included in both of these calculations likely inflate the number of bits the measurement itself wastes. We cannot deconvolve only the preparation error from the measurement results, but we can zero the contribution of preparation to $\kappa t^{(0)}$. We emphasize that $\kappa t^{(0)}$ will always contain a contribution from the previous experimental step, but we can construct a $C$ confined to errors that solely occur during the measurement duration. This scenario yields $4-\langle S_i\rangle_{\text{C, no prep}}=3.17$ bits, which is likely a lower bound as the residuals in Fig.~\ref{fig3} show that $C$ does not perfectly capture all errors.

Finally, we can use this method to estimate the single-shot error budget. With $\kappa t^{(k)}$ as in Table \ref{measured_error_parameters} and unit fidelity $\text{C}\hat{\mathcal{P}}^k$ pulses and ancilla readout ($\epsilon_{g}^{(k)}=\epsilon_{e}^{(k)}=0$) we find $4-\langle S_i\rangle_{\text{only }\kappa t^{(k)}}=3.72$ bits. The remaining information is lost to photon loss in the storage mode. This suggests that significant gains can be made by optimizing the $\text{C}\hat{\mathcal{P}}^k$ pulses and ancilla readout before needing to decrease $\kappa t^{(k)}$.

\begin{figure}
	\centering
	\includegraphics[scale=1]{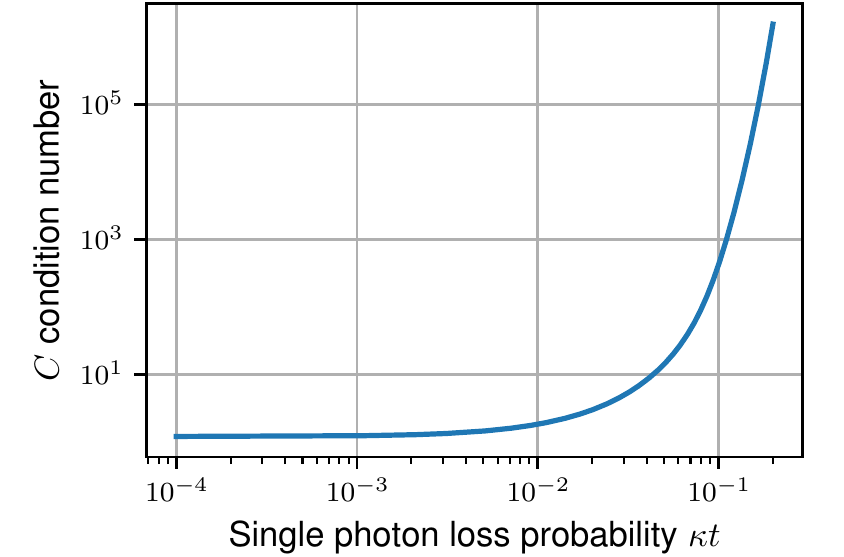}
	\caption{\label{supp_fig3} Condition number vs. storage mode error rate. As the cavity error rate increases, the condition number (ratio of largest to smallest singular value) of $C$ diverges, indicating that the inversion of $C$ is no longer numerically stable. For simplicity, we choose typical experimental values for the errors $\epsilon_{g}^{(k)}=0.01,\,\epsilon_{e}^{(k)}=0.03$ and sweep all $\kappa t^{(k)}=\kappa t'$ simultaneously. } 
\end{figure}

\section{The invertibility of $C$}
\label{C_invertibility}
Our error mitigation protocol works in the range of error rates that allow robust inversion of $C$. In the limit of small error rates, $C$ approaches the identity. As the error rates $\epsilon_{g}^{(k)}, \epsilon_{e}^{(k)}$, and $\kappa t^{(k)}$ become large, we expect that $C$ will become ill-conditioned. For example, as the storage mode error rates $\kappa t^{(k)}$ approach $1$, the probability $C_{0n}$ of any photon state $\ket{n}$ decaying to $\ket{0}$ approaches 1. But since the column vectors of $C$ are normalized to 1, they approach degeneracy and prevent the inversion of $C$. This exact scenario is illustrated in Fig.~\ref{supp_fig3}. Similarly, the condition number also diverges as $\epsilon_{g}^{(k)}, \epsilon_{e}^{(k)}\rightarrow1$. We operate the small error rate regime in which $C$ is robustly invertible.

\section{Confusion matrix expansion}
\label{confusion matrix expansion appendix}
If simply correcting peak intensities in a measured spectrum is insufficient, we can expand the confusion matrix
about the identity matrix. By truncating this expansion to some order
so that the total number of relevant matrix elements is polynomial,
we can calculate polynomially many of the largest entries in $\mathbf{P}^{\text{ideal}}$.
Using this approach, we may find significant peaks $\mathbf{P}_{s}^{\text{ideal}}\neq0$
even when $\mathbf{P}_{s}^{\text{meas}}=0$. However, each peak height will
only be accurate to the order at which the expansion is truncated.

Consider the confusion matrix $C_{m}$ for photon number readout of
the $m^{\text{th}}$ mode. Let us assume that $C_{m}$ is close to the identity,
so that its inverse is also close to the identity,
\begin{equation}
\label{C_inv_expansion}
C_{m}^{-1}=I+\varepsilon\mathcal{E}_{m}
\end{equation}
where $\varepsilon\ll1$ is some small number, and $\mathcal{E}_{m}$
is a matrix normalized so that
\begin{equation}
\left\Vert \mathcal{E}_{m}\right\Vert _{\max}=\max_{ij}\left|\left[\mathcal{E}_{m}\right]_{ij}\right|=1.
\end{equation}
Thus,
\begin{equation}
\mathbf{P}^{\text{mit}}=\left[\left(I+\varepsilon\mathcal{E}_{1}\right)\otimes\left(I+\varepsilon\mathcal{E}_{2}\right)\otimes\ldots\otimes\left(I+\varepsilon\mathcal{E}_{M}\right)\right]\mathbf{P}^{\text{meas}}
\end{equation}
Now, the main idea is to use this expansion to compute only the entries
of $\mathbf{P}^{\text{mit}}$ which are $O(\varepsilon^{q})$, then neglect all
entries which go as higher powers in $\varepsilon$. To perform this
computation, first notice that we only need to consider a small subset
of the columns of $C^{-1}$ because $\mathbf{P}^{\text{meas}}$ is sparse.
That is, only the columns $\left[C^{-1}\right]_{*,j\in\mathcal{S}}$
contribute, where $\left[C^{-1}\right]_{*,j}$ denotes the $j^{\text{th}}$
column of $C^{-1}$. While each such column contains exponentially
many nonzero entries $N_{\text{max}}^M$ in general, it turns out that the columns may
contain only polynomially many nonzero entries when expanded in $\varepsilon$,
depending on the order of the expansion. To compute $\mathbf{P}^{\text{mit}}$
we then compute all entries which are nonzero to order $q$ in the
columns $\left[C^{-1}\right]_{*,j\in\mathcal{S}}$, giving us a sparse
representation of $C^{-1}$. It is then efficient
to multiply $C^{-1}\mathbf{P}^{\text{meas}}$.

Each entry in $C^{-1}$ is a product of entries from the single-mode confusion matrices $C_{k}^{-1}$. In each column of $C_{k}^{-1}$, there
are $N_{\text{max}}$ elements, one of which is $O(1)$ and up to $N_{\text{max}}-1$ of which are
$O(\varepsilon)$ as defined in Eq. \ref{C_inv_expansion}. Each of the $N_{\text{max}}^M$ elements in a column of $C^{-1}$ is a product of $M$ elements where the $k^{\text{th}}$ element in the product is drawn from a fixed column of $C^{-1}_k$. Our task is to count the number such products that result in an element of $O(\varepsilon^q)$. We choose $q$ columns out of $M$ to each contribute a factor of $\varepsilon$, and within each column we choose one of $N_{\text{max}}-1$ entries that are $O(\varepsilon)$. There are thus at most $\binom{M}{q}\left(N_{\text{max}}-1\right)^q$ entries in a column of $C^{-1}$ which are polynomials of degree $q$ in $\varepsilon$. Therefore, the total number of nonzero entries
$\mathcal{N}$ per column which must be calculated when expanding $\left[C^{-1}\right]_{*,j\in\mathcal{S}}$
to $q^{\text{th}}$ order in $\varepsilon$ is 
\[
\mathcal{N}=\sum_{\ell=0}^{q}(N_{\text{max}}-1)^{\ell}\binom{M}{\ell}.
\]
If $\mathcal{N}$ is only polynomially large, this approach is efficient, as the sparsity of $\mathbf{P}^{\text{meas}}$ ensures that the total number of entries $\mathcal{N}|\mathcal{S}|$ in $\left[C^{-1}\right]_{*,j\in\mathcal{S}}$ is not exponentially large.

We now examine the scaling of $\mathcal{N}$, which can be written as $\mathcal{N}=O(N_{\text{max}}^{\ell})\times\sum_{\ell=0}^{q}\binom{M}{\ell}$. The partial sum of binomial coefficients $\sum_{\ell=0}^{q}\binom{M}{\ell}$
does not have a closed form expression; for small values of $q$ it
is $O(M^{q})$, but for $q=M$ it is $O(2^{M})$ . Thus, whether this
approach is efficient depends on the choice of $q$. For small $q$, we have $\mathcal{N}=O\left(N_{\text{max}}^{q}M^{q}\right)$ so that the total
number of nonzero entries scales polynomially in the number of modes
$M$ and the maximal photon number $N_{\text{max}}$, ensuring the efficiency of the expansion.

\bibliography{sources}

\begin{thebibliography}{68}%
\makeatletter
\providecommand \@ifxundefined [1]{%
 \@ifx{#1\undefined}
}%
\providecommand \@ifnum [1]{%
 \ifnum #1\expandafter \@firstoftwo
 \else \expandafter \@secondoftwo
 \fi
}%
\providecommand \@ifx [1]{%
 \ifx #1\expandafter \@firstoftwo
 \else \expandafter \@secondoftwo
 \fi
}%
\providecommand \natexlab [1]{#1}%
\providecommand \enquote  [1]{``#1''}%
\providecommand \bibnamefont  [1]{#1}%
\providecommand \bibfnamefont [1]{#1}%
\providecommand \citenamefont [1]{#1}%
\providecommand \href@noop [0]{\@secondoftwo}%
\providecommand \href [0]{\begingroup \@sanitize@url \@href}%
\providecommand \@href[1]{\@@startlink{#1}\@@href}%
\providecommand \@@href[1]{\endgroup#1\@@endlink}%
\providecommand \@sanitize@url [0]{\catcode `\\12\catcode `\$12\catcode
  `\&12\catcode `\#12\catcode `\^12\catcode `\_12\catcode `\%12\relax}%
\providecommand \@@startlink[1]{}%
\providecommand \@@endlink[0]{}%
\providecommand \url  [0]{\begingroup\@sanitize@url \@url }%
\providecommand \@url [1]{\endgroup\@href {#1}{\urlprefix }}%
\providecommand \urlprefix  [0]{URL }%
\providecommand \Eprint [0]{\href }%
\providecommand \doibase [0]{https://doi.org/}%
\providecommand \selectlanguage [0]{\@gobble}%
\providecommand \bibinfo  [0]{\@secondoftwo}%
\providecommand \bibfield  [0]{\@secondoftwo}%
\providecommand \translation [1]{[#1]}%
\providecommand \BibitemOpen [0]{}%
\providecommand \bibitemStop [0]{}%
\providecommand \bibitemNoStop [0]{.\EOS\space}%
\providecommand \EOS [0]{\spacefactor3000\relax}%
\providecommand \BibitemShut  [1]{\csname bibitem#1\endcsname}%
\let\auto@bib@innerbib\@empty
%</preamble>
\bibitem [{\citenamefont {Lo}\ \emph {et~al.}(2012)\citenamefont {Lo},
  \citenamefont {Curty},\ and\ \citenamefont
  {Qi}}]{lo_measurement-device-independent_2012}%
  \BibitemOpen
  \bibfield  {author} {\bibinfo {author} {\bibfnamefont {H.-K.}\ \bibnamefont
  {Lo}}, \bibinfo {author} {\bibfnamefont {M.}~\bibnamefont {Curty}},\ and\
  \bibinfo {author} {\bibfnamefont {B.}~\bibnamefont {Qi}},\ }\bibfield
  {title} {\bibinfo {title} {Measurement-{Device}-{Independent} {Quantum} {Key}
  {Distribution}},\ }\href {https://doi.org/10.1103/PhysRevLett.108.130503}
  {\bibfield  {journal} {\bibinfo  {journal} {Phys. Rev. Lett.}\ }\textbf
  {\bibinfo {volume} {108}},\ \bibinfo {pages} {130503} (\bibinfo {year}
  {2012})}\BibitemShut {NoStop}%
\bibitem [{\citenamefont {Yin}\ \emph {et~al.}(2016)\citenamefont {Yin},
  \citenamefont {Chen}, \citenamefont {Yu}, \citenamefont {Liu}, \citenamefont
  {You}, \citenamefont {Zhou}, \citenamefont {Chen}, \citenamefont {Mao},
  \citenamefont {Huang}, \citenamefont {Zhang}, \citenamefont {Chen},
  \citenamefont {Li}, \citenamefont {Nolan}, \citenamefont {Zhou},
  \citenamefont {Jiang}, \citenamefont {Wang}, \citenamefont {Zhang},
  \citenamefont {Wang},\ and\ \citenamefont
  {Pan}}]{yin_measurement-device-independent_2016}%
  \BibitemOpen
  \bibfield  {author} {\bibinfo {author} {\bibfnamefont {H.-L.}\ \bibnamefont
  {Yin}}, \bibinfo {author} {\bibfnamefont {T.-Y.}\ \bibnamefont {Chen}},
  \bibinfo {author} {\bibfnamefont {Z.-W.}\ \bibnamefont {Yu}}, \bibinfo
  {author} {\bibfnamefont {H.}~\bibnamefont {Liu}}, \bibinfo {author}
  {\bibfnamefont {L.-X.}\ \bibnamefont {You}}, \bibinfo {author} {\bibfnamefont
  {Y.-H.}\ \bibnamefont {Zhou}}, \bibinfo {author} {\bibfnamefont {S.-J.}\
  \bibnamefont {Chen}}, \bibinfo {author} {\bibfnamefont {Y.}~\bibnamefont
  {Mao}}, \bibinfo {author} {\bibfnamefont {M.-Q.}\ \bibnamefont {Huang}},
  \bibinfo {author} {\bibfnamefont {W.-J.}\ \bibnamefont {Zhang}}, \bibinfo
  {author} {\bibfnamefont {H.}~\bibnamefont {Chen}}, \bibinfo {author}
  {\bibfnamefont {M.~J.}\ \bibnamefont {Li}}, \bibinfo {author} {\bibfnamefont
  {D.}~\bibnamefont {Nolan}}, \bibinfo {author} {\bibfnamefont
  {F.}~\bibnamefont {Zhou}}, \bibinfo {author} {\bibfnamefont {X.}~\bibnamefont
  {Jiang}}, \bibinfo {author} {\bibfnamefont {Z.}~\bibnamefont {Wang}},
  \bibinfo {author} {\bibfnamefont {Q.}~\bibnamefont {Zhang}}, \bibinfo
  {author} {\bibfnamefont {X.-B.}\ \bibnamefont {Wang}},\ and\ \bibinfo
  {author} {\bibfnamefont {J.-W.}\ \bibnamefont {Pan}},\ }\bibfield  {title}
  {\bibinfo {title} {Measurement-{Device}-{Independent} {Quantum} {Key}
  {Distribution} {Over} a 404 km {Optical} {Fiber}},\ }\href
  {https://doi.org/10.1103/PhysRevLett.117.190501} {\bibfield  {journal}
  {\bibinfo  {journal} {Phys. Rev. Lett.}\ }\textbf {\bibinfo {volume} {117}},\
  \bibinfo {pages} {190501} (\bibinfo {year} {2016})}\BibitemShut {NoStop}%
\bibitem [{\citenamefont {Liao}\ \emph {et~al.}(2017)\citenamefont {Liao},
  \citenamefont {Cai}, \citenamefont {Liu}, \citenamefont {Zhang},
  \citenamefont {Li}, \citenamefont {Ren}, \citenamefont {Yin}, \citenamefont
  {Shen}, \citenamefont {Cao}, \citenamefont {Li}, \citenamefont {Li},
  \citenamefont {Chen}, \citenamefont {Sun}, \citenamefont {Jia}, \citenamefont
  {Wu}, \citenamefont {Jiang}, \citenamefont {Wang}, \citenamefont {Huang},
  \citenamefont {Wang}, \citenamefont {Zhou}, \citenamefont {Deng},
  \citenamefont {Xi}, \citenamefont {Ma}, \citenamefont {Hu}, \citenamefont
  {Zhang}, \citenamefont {Chen}, \citenamefont {Liu}, \citenamefont {Wang},
  \citenamefont {Zhu}, \citenamefont {Lu}, \citenamefont {Shu}, \citenamefont
  {Peng}, \citenamefont {Wang},\ and\ \citenamefont
  {Pan}}]{liao_satellite--ground_2017}%
  \BibitemOpen
  \bibfield  {author} {\bibinfo {author} {\bibfnamefont {S.-K.}\ \bibnamefont
  {Liao}}, \bibinfo {author} {\bibfnamefont {W.-Q.}\ \bibnamefont {Cai}},
  \bibinfo {author} {\bibfnamefont {W.-Y.}\ \bibnamefont {Liu}}, \bibinfo
  {author} {\bibfnamefont {L.}~\bibnamefont {Zhang}}, \bibinfo {author}
  {\bibfnamefont {Y.}~\bibnamefont {Li}}, \bibinfo {author} {\bibfnamefont
  {J.-G.}\ \bibnamefont {Ren}}, \bibinfo {author} {\bibfnamefont
  {J.}~\bibnamefont {Yin}}, \bibinfo {author} {\bibfnamefont {Q.}~\bibnamefont
  {Shen}}, \bibinfo {author} {\bibfnamefont {Y.}~\bibnamefont {Cao}}, \bibinfo
  {author} {\bibfnamefont {Z.-P.}\ \bibnamefont {Li}}, \bibinfo {author}
  {\bibfnamefont {F.-Z.}\ \bibnamefont {Li}}, \bibinfo {author} {\bibfnamefont
  {X.-W.}\ \bibnamefont {Chen}}, \bibinfo {author} {\bibfnamefont {L.-H.}\
  \bibnamefont {Sun}}, \bibinfo {author} {\bibfnamefont {J.-J.}\ \bibnamefont
  {Jia}}, \bibinfo {author} {\bibfnamefont {J.-C.}\ \bibnamefont {Wu}},
  \bibinfo {author} {\bibfnamefont {X.-J.}\ \bibnamefont {Jiang}}, \bibinfo
  {author} {\bibfnamefont {J.-F.}\ \bibnamefont {Wang}}, \bibinfo {author}
  {\bibfnamefont {Y.-M.}\ \bibnamefont {Huang}}, \bibinfo {author}
  {\bibfnamefont {Q.}~\bibnamefont {Wang}}, \bibinfo {author} {\bibfnamefont
  {Y.-L.}\ \bibnamefont {Zhou}}, \bibinfo {author} {\bibfnamefont
  {L.}~\bibnamefont {Deng}}, \bibinfo {author} {\bibfnamefont {T.}~\bibnamefont
  {Xi}}, \bibinfo {author} {\bibfnamefont {L.}~\bibnamefont {Ma}}, \bibinfo
  {author} {\bibfnamefont {T.}~\bibnamefont {Hu}}, \bibinfo {author}
  {\bibfnamefont {Q.}~\bibnamefont {Zhang}}, \bibinfo {author} {\bibfnamefont
  {Y.-A.}\ \bibnamefont {Chen}}, \bibinfo {author} {\bibfnamefont {N.-L.}\
  \bibnamefont {Liu}}, \bibinfo {author} {\bibfnamefont {X.-B.}\ \bibnamefont
  {Wang}}, \bibinfo {author} {\bibfnamefont {Z.-C.}\ \bibnamefont {Zhu}},
  \bibinfo {author} {\bibfnamefont {C.-Y.}\ \bibnamefont {Lu}}, \bibinfo
  {author} {\bibfnamefont {R.}~\bibnamefont {Shu}}, \bibinfo {author}
  {\bibfnamefont {C.-Z.}\ \bibnamefont {Peng}}, \bibinfo {author}
  {\bibfnamefont {J.-Y.}\ \bibnamefont {Wang}},\ and\ \bibinfo {author}
  {\bibfnamefont {J.-W.}\ \bibnamefont {Pan}},\ }\bibfield  {title} {\bibinfo
  {title} {Satellite-to-ground quantum key distribution},\ }\bibfield
  {journal} {\bibinfo  {journal} {Nature}\ }\textbf {\bibinfo {volume} {549}},\
  \href {https://doi.org/10.1038/nature23655} {10.1038/nature23655} (\bibinfo
  {year} {2017})\BibitemShut {NoStop}%
\bibitem [{\citenamefont {Knill}\ \emph {et~al.}(2001)\citenamefont {Knill},
  \citenamefont {Laflamme},\ and\ \citenamefont {Milburn}}]{knill_scheme_2001}%
  \BibitemOpen
  \bibfield  {author} {\bibinfo {author} {\bibfnamefont {E.}~\bibnamefont
  {Knill}}, \bibinfo {author} {\bibfnamefont {R.}~\bibnamefont {Laflamme}},\
  and\ \bibinfo {author} {\bibfnamefont {G.~J.}\ \bibnamefont {Milburn}},\
  }\bibfield  {title} {\bibinfo {title} {A scheme for efficient quantum
  computation with linear optics},\ }\bibfield  {journal} {\bibinfo  {journal}
  {Nature}\ }\textbf {\bibinfo {volume} {409}},\ \href
  {https://doi.org/10.1038/35051009} {10.1038/35051009} (\bibinfo {year}
  {2001})\BibitemShut {NoStop}%
\bibitem [{\citenamefont {Kok}\ \emph {et~al.}(2007)\citenamefont {Kok},
  \citenamefont {Munro}, \citenamefont {Nemoto}, \citenamefont {Ralph},
  \citenamefont {Dowling},\ and\ \citenamefont {Milburn}}]{kok_linear_2007}%
  \BibitemOpen
  \bibfield  {author} {\bibinfo {author} {\bibfnamefont {P.}~\bibnamefont
  {Kok}}, \bibinfo {author} {\bibfnamefont {W.~J.}\ \bibnamefont {Munro}},
  \bibinfo {author} {\bibfnamefont {K.}~\bibnamefont {Nemoto}}, \bibinfo
  {author} {\bibfnamefont {T.~C.}\ \bibnamefont {Ralph}}, \bibinfo {author}
  {\bibfnamefont {J.~P.}\ \bibnamefont {Dowling}},\ and\ \bibinfo {author}
  {\bibfnamefont {G.~J.}\ \bibnamefont {Milburn}},\ }\bibfield  {title}
  {\bibinfo {title} {Linear optical quantum computing with photonic qubits},\
  }\href {https://doi.org/10.1103/RevModPhys.79.135} {\bibfield  {journal}
  {\bibinfo  {journal} {Rev. Mod. Phys.}\ }\textbf {\bibinfo {volume} {79}},\
  \bibinfo {pages} {135} (\bibinfo {year} {2007})}\BibitemShut {NoStop}%
\bibitem [{\citenamefont {Duan}\ \emph {et~al.}(2001)\citenamefont {Duan},
  \citenamefont {Lukin}, \citenamefont {Cirac},\ and\ \citenamefont
  {Zoller}}]{duan_long-distance_2001}%
  \BibitemOpen
  \bibfield  {author} {\bibinfo {author} {\bibfnamefont {L.-M.}\ \bibnamefont
  {Duan}}, \bibinfo {author} {\bibfnamefont {M.~D.}\ \bibnamefont {Lukin}},
  \bibinfo {author} {\bibfnamefont {J.~I.}\ \bibnamefont {Cirac}},\ and\
  \bibinfo {author} {\bibfnamefont {P.}~\bibnamefont {Zoller}},\ }\bibfield
  {title} {\bibinfo {title} {Long-distance quantum communication with atomic
  ensembles and linear optics},\ }\bibfield  {journal} {\bibinfo  {journal}
  {Nature}\ }\textbf {\bibinfo {volume} {414}},\ \href
  {https://doi.org/10.1038/35106500} {10.1038/35106500} (\bibinfo {year}
  {2001})\BibitemShut {NoStop}%
\bibitem [{\citenamefont {Ursin}\ \emph {et~al.}(2007)\citenamefont {Ursin},
  \citenamefont {Tiefenbacher}, \citenamefont {Schmitt-Manderbach},
  \citenamefont {Weier}, \citenamefont {Scheidl}, \citenamefont {Lindenthal},
  \citenamefont {Blauensteiner}, \citenamefont {Jennewein}, \citenamefont
  {Perdigues}, \citenamefont {Trojek}, \citenamefont {Ömer}, \citenamefont
  {Fürst}, \citenamefont {Meyenburg}, \citenamefont {Rarity}, \citenamefont
  {Sodnik}, \citenamefont {Barbieri}, \citenamefont {Weinfurter},\ and\
  \citenamefont {Zeilinger}}]{ursin_entanglement-based_2007}%
  \BibitemOpen
  \bibfield  {author} {\bibinfo {author} {\bibfnamefont {R.}~\bibnamefont
  {Ursin}}, \bibinfo {author} {\bibfnamefont {F.}~\bibnamefont {Tiefenbacher}},
  \bibinfo {author} {\bibfnamefont {T.}~\bibnamefont {Schmitt-Manderbach}},
  \bibinfo {author} {\bibfnamefont {H.}~\bibnamefont {Weier}}, \bibinfo
  {author} {\bibfnamefont {T.}~\bibnamefont {Scheidl}}, \bibinfo {author}
  {\bibfnamefont {M.}~\bibnamefont {Lindenthal}}, \bibinfo {author}
  {\bibfnamefont {B.}~\bibnamefont {Blauensteiner}}, \bibinfo {author}
  {\bibfnamefont {T.}~\bibnamefont {Jennewein}}, \bibinfo {author}
  {\bibfnamefont {J.}~\bibnamefont {Perdigues}}, \bibinfo {author}
  {\bibfnamefont {P.}~\bibnamefont {Trojek}}, \bibinfo {author} {\bibfnamefont
  {B.}~\bibnamefont {Ömer}}, \bibinfo {author} {\bibfnamefont
  {M.}~\bibnamefont {Fürst}}, \bibinfo {author} {\bibfnamefont
  {M.}~\bibnamefont {Meyenburg}}, \bibinfo {author} {\bibfnamefont
  {J.}~\bibnamefont {Rarity}}, \bibinfo {author} {\bibfnamefont
  {Z.}~\bibnamefont {Sodnik}}, \bibinfo {author} {\bibfnamefont
  {C.}~\bibnamefont {Barbieri}}, \bibinfo {author} {\bibfnamefont
  {H.}~\bibnamefont {Weinfurter}},\ and\ \bibinfo {author} {\bibfnamefont
  {A.}~\bibnamefont {Zeilinger}},\ }\bibfield  {title} {\bibinfo {title}
  {Entanglement-based quantum communication over 144 km},\ }\bibfield
  {journal} {\bibinfo  {journal} {Nature Physics}\ }\textbf {\bibinfo {volume}
  {3}},\ \href {https://doi.org/10.1038/nphys629} {10.1038/nphys629} (\bibinfo
  {year} {2007})\BibitemShut {NoStop}%
\bibitem [{\citenamefont {Kimble}(2008)}]{kimble_quantum_2008}%
  \BibitemOpen
  \bibfield  {author} {\bibinfo {author} {\bibfnamefont {H.~J.}\ \bibnamefont
  {Kimble}},\ }\bibfield  {title} {\bibinfo {title} {The quantum internet},\
  }\bibfield  {journal} {\bibinfo  {journal} {Nature}\ }\textbf {\bibinfo
  {volume} {453}},\ \href {https://doi.org/10.1038/nature07127}
  {10.1038/nature07127} (\bibinfo {year} {2008})\BibitemShut {NoStop}%
\bibitem [{\citenamefont {Northup}\ and\ \citenamefont
  {Blatt}(2014)}]{northup_quantum_2014}%
  \BibitemOpen
  \bibfield  {author} {\bibinfo {author} {\bibfnamefont {T.~E.}\ \bibnamefont
  {Northup}}\ and\ \bibinfo {author} {\bibfnamefont {R.}~\bibnamefont
  {Blatt}},\ }\bibfield  {title} {\bibinfo {title} {Quantum information
  transfer using photons},\ }\bibfield  {journal} {\bibinfo  {journal} {Nature
  Photonics}\ }\textbf {\bibinfo {volume} {8}},\ \href
  {https://doi.org/10.1038/nphoton.2014.53} {10.1038/nphoton.2014.53} (\bibinfo
  {year} {2014})\BibitemShut {NoStop}%
\bibitem [{\citenamefont {Hu}\ \emph {et~al.}(2016)\citenamefont {Hu},
  \citenamefont {Yu}, \citenamefont {Jing}, \citenamefont {Xiao}, \citenamefont
  {Jia}, \citenamefont {Qin},\ and\ \citenamefont
  {Long}}]{hu_experimental_2016}%
  \BibitemOpen
  \bibfield  {author} {\bibinfo {author} {\bibfnamefont {J.-Y.}\ \bibnamefont
  {Hu}}, \bibinfo {author} {\bibfnamefont {B.}~\bibnamefont {Yu}}, \bibinfo
  {author} {\bibfnamefont {M.-Y.}\ \bibnamefont {Jing}}, \bibinfo {author}
  {\bibfnamefont {L.-T.}\ \bibnamefont {Xiao}}, \bibinfo {author}
  {\bibfnamefont {S.-T.}\ \bibnamefont {Jia}}, \bibinfo {author} {\bibfnamefont
  {G.-Q.}\ \bibnamefont {Qin}},\ and\ \bibinfo {author} {\bibfnamefont {G.-L.}\
  \bibnamefont {Long}},\ }\bibfield  {title} {\bibinfo {title} {Experimental
  quantum secure direct communication with single photons},\ }\bibfield
  {journal} {\bibinfo  {journal} {Light: Science \& Applications}\ }\textbf
  {\bibinfo {volume} {5}},\ \href {https://doi.org/10.1038/lsa.2016.144}
  {10.1038/lsa.2016.144} (\bibinfo {year} {2016})\BibitemShut {NoStop}%
\bibitem [{\citenamefont {Zhang}\ \emph {et~al.}(2017)\citenamefont {Zhang},
  \citenamefont {Ding}, \citenamefont {Sheng}, \citenamefont {Zhou},
  \citenamefont {Shi},\ and\ \citenamefont {Guo}}]{zhang_quantum_2017}%
  \BibitemOpen
  \bibfield  {author} {\bibinfo {author} {\bibfnamefont {W.}~\bibnamefont
  {Zhang}}, \bibinfo {author} {\bibfnamefont {D.-S.}\ \bibnamefont {Ding}},
  \bibinfo {author} {\bibfnamefont {Y.-B.}\ \bibnamefont {Sheng}}, \bibinfo
  {author} {\bibfnamefont {L.}~\bibnamefont {Zhou}}, \bibinfo {author}
  {\bibfnamefont {B.-S.}\ \bibnamefont {Shi}},\ and\ \bibinfo {author}
  {\bibfnamefont {G.-C.}\ \bibnamefont {Guo}},\ }\bibfield  {title} {\bibinfo
  {title} {Quantum {Secure} {Direct} {Communication} with {Quantum} {Memory}},\
  }\href {https://doi.org/10.1103/PhysRevLett.118.220501} {\bibfield  {journal}
  {\bibinfo  {journal} {Phys. Rev. Lett.}\ }\textbf {\bibinfo {volume} {118}},\
  \bibinfo {pages} {220501} (\bibinfo {year} {2017})}\BibitemShut {NoStop}%
\bibitem [{\citenamefont {Dou}\ \emph {et~al.}(2018)\citenamefont {Dou},
  \citenamefont {Yang}, \citenamefont {Du}, \citenamefont {Lao}, \citenamefont
  {Gao}, \citenamefont {Qiao}, \citenamefont {Li}, \citenamefont {Pang},
  \citenamefont {Feng}, \citenamefont {Tang},\ and\ \citenamefont
  {Jin}}]{dou_broadband_2018}%
  \BibitemOpen
  \bibfield  {author} {\bibinfo {author} {\bibfnamefont {J.-P.}\ \bibnamefont
  {Dou}}, \bibinfo {author} {\bibfnamefont {A.-L.}\ \bibnamefont {Yang}},
  \bibinfo {author} {\bibfnamefont {M.-Y.}\ \bibnamefont {Du}}, \bibinfo
  {author} {\bibfnamefont {D.}~\bibnamefont {Lao}}, \bibinfo {author}
  {\bibfnamefont {J.}~\bibnamefont {Gao}}, \bibinfo {author} {\bibfnamefont
  {L.-F.}\ \bibnamefont {Qiao}}, \bibinfo {author} {\bibfnamefont
  {H.}~\bibnamefont {Li}}, \bibinfo {author} {\bibfnamefont {X.-L.}\
  \bibnamefont {Pang}}, \bibinfo {author} {\bibfnamefont {Z.}~\bibnamefont
  {Feng}}, \bibinfo {author} {\bibfnamefont {H.}~\bibnamefont {Tang}},\ and\
  \bibinfo {author} {\bibfnamefont {X.-M.}\ \bibnamefont {Jin}},\ }\bibfield
  {title} {\bibinfo {title} {A broadband {DLCZ} quantum memory in
  room-temperature atoms},\ }\bibfield  {journal} {\bibinfo  {journal}
  {Communications Physics}\ }\textbf {\bibinfo {volume} {1}},\ \href
  {https://doi.org/10.1038/s42005-018-0057-9} {10.1038/s42005-018-0057-9}
  (\bibinfo {year} {2018})\BibitemShut {NoStop}%
\bibitem [{\citenamefont {Spring}\ \emph {et~al.}(2013)\citenamefont {Spring},
  \citenamefont {Metcalf}, \citenamefont {Humphreys}, \citenamefont
  {Kolthammer}, \citenamefont {Jin}, \citenamefont {Barbieri}, \citenamefont
  {Datta}, \citenamefont {Thomas-Peter}, \citenamefont {Langford},
  \citenamefont {Kundys}, \citenamefont {Gates}, \citenamefont {Smith},
  \citenamefont {Smith},\ and\ \citenamefont {Walmsley}}]{spring_boson_2013}%
  \BibitemOpen
  \bibfield  {author} {\bibinfo {author} {\bibfnamefont {J.~B.}\ \bibnamefont
  {Spring}}, \bibinfo {author} {\bibfnamefont {B.~J.}\ \bibnamefont {Metcalf}},
  \bibinfo {author} {\bibfnamefont {P.~C.}\ \bibnamefont {Humphreys}}, \bibinfo
  {author} {\bibfnamefont {W.~S.}\ \bibnamefont {Kolthammer}}, \bibinfo
  {author} {\bibfnamefont {X.-M.}\ \bibnamefont {Jin}}, \bibinfo {author}
  {\bibfnamefont {M.}~\bibnamefont {Barbieri}}, \bibinfo {author}
  {\bibfnamefont {A.}~\bibnamefont {Datta}}, \bibinfo {author} {\bibfnamefont
  {N.}~\bibnamefont {Thomas-Peter}}, \bibinfo {author} {\bibfnamefont {N.~K.}\
  \bibnamefont {Langford}}, \bibinfo {author} {\bibfnamefont {D.}~\bibnamefont
  {Kundys}}, \bibinfo {author} {\bibfnamefont {J.~C.}\ \bibnamefont {Gates}},
  \bibinfo {author} {\bibfnamefont {B.~J.}\ \bibnamefont {Smith}}, \bibinfo
  {author} {\bibfnamefont {P.~G.~R.}\ \bibnamefont {Smith}},\ and\ \bibinfo
  {author} {\bibfnamefont {I.~A.}\ \bibnamefont {Walmsley}},\ }\bibfield
  {title} {\bibinfo {title} {Boson {Sampling} on a {Photonic} {Chip}},\ }\href
  {https://doi.org/10.1126/science.1231692} {\bibfield  {journal} {\bibinfo
  {journal} {Science}\ }\textbf {\bibinfo {volume} {339}},\ \bibinfo {pages}
  {798} (\bibinfo {year} {2013})}\BibitemShut {NoStop}%
\bibitem [{\citenamefont {Tillmann}\ \emph {et~al.}(2013)\citenamefont
  {Tillmann}, \citenamefont {Dakić}, \citenamefont {Heilmann}, \citenamefont
  {Nolte}, \citenamefont {Szameit},\ and\ \citenamefont
  {Walther}}]{tillmann_experimental_2013}%
  \BibitemOpen
  \bibfield  {author} {\bibinfo {author} {\bibfnamefont {M.}~\bibnamefont
  {Tillmann}}, \bibinfo {author} {\bibfnamefont {B.}~\bibnamefont {Dakić}},
  \bibinfo {author} {\bibfnamefont {R.}~\bibnamefont {Heilmann}}, \bibinfo
  {author} {\bibfnamefont {S.}~\bibnamefont {Nolte}}, \bibinfo {author}
  {\bibfnamefont {A.}~\bibnamefont {Szameit}},\ and\ \bibinfo {author}
  {\bibfnamefont {P.}~\bibnamefont {Walther}},\ }\bibfield  {title} {\bibinfo
  {title} {Experimental boson sampling},\ }\bibfield  {journal} {\bibinfo
  {journal} {Nature Photonics}\ }\textbf {\bibinfo {volume} {7}},\ \href
  {https://doi.org/10.1038/nphoton.2013.102} {10.1038/nphoton.2013.102}
  (\bibinfo {year} {2013})\BibitemShut {NoStop}%
\bibitem [{\citenamefont {Sparrow}\ \emph {et~al.}(2018)\citenamefont
  {Sparrow}, \citenamefont {Martín-López}, \citenamefont {Maraviglia},
  \citenamefont {Neville}, \citenamefont {Harrold}, \citenamefont {Carolan},
  \citenamefont {Joglekar}, \citenamefont {Hashimoto}, \citenamefont {Matsuda},
  \citenamefont {O’Brien}, \citenamefont {Tew},\ and\ \citenamefont
  {Laing}}]{sparrow_simulating_2018}%
  \BibitemOpen
  \bibfield  {author} {\bibinfo {author} {\bibfnamefont {C.}~\bibnamefont
  {Sparrow}}, \bibinfo {author} {\bibfnamefont {E.}~\bibnamefont
  {Martín-López}}, \bibinfo {author} {\bibfnamefont {N.}~\bibnamefont
  {Maraviglia}}, \bibinfo {author} {\bibfnamefont {A.}~\bibnamefont {Neville}},
  \bibinfo {author} {\bibfnamefont {C.}~\bibnamefont {Harrold}}, \bibinfo
  {author} {\bibfnamefont {J.}~\bibnamefont {Carolan}}, \bibinfo {author}
  {\bibfnamefont {Y.~N.}\ \bibnamefont {Joglekar}}, \bibinfo {author}
  {\bibfnamefont {T.}~\bibnamefont {Hashimoto}}, \bibinfo {author}
  {\bibfnamefont {N.}~\bibnamefont {Matsuda}}, \bibinfo {author} {\bibfnamefont
  {J.~L.}\ \bibnamefont {O’Brien}}, \bibinfo {author} {\bibfnamefont {D.~P.}\
  \bibnamefont {Tew}},\ and\ \bibinfo {author} {\bibfnamefont {A.}~\bibnamefont
  {Laing}},\ }\bibfield  {title} {\bibinfo {title} {Simulating the vibrational
  quantum dynamics of molecules using photonics},\ }\bibfield  {journal}
  {\bibinfo  {journal} {Nature}\ }\textbf {\bibinfo {volume} {557}},\ \href
  {https://doi.org/10.1038/s41586-018-0152-9} {10.1038/s41586-018-0152-9}
  (\bibinfo {year} {2018})\BibitemShut {NoStop}%
\bibitem [{\citenamefont {Bentivegna}\ \emph {et~al.}(2015)\citenamefont
  {Bentivegna}, \citenamefont {Spagnolo}, \citenamefont {Vitelli},
  \citenamefont {Flamini}, \citenamefont {Viggianiello}, \citenamefont
  {Latmiral}, \citenamefont {Mataloni}, \citenamefont {Brod}, \citenamefont
  {Galvão}, \citenamefont {Crespi}, \citenamefont {Ramponi}, \citenamefont
  {Osellame},\ and\ \citenamefont {Sciarrino}}]{bentivegna_experimental_2015}%
  \BibitemOpen
  \bibfield  {author} {\bibinfo {author} {\bibfnamefont {M.}~\bibnamefont
  {Bentivegna}}, \bibinfo {author} {\bibfnamefont {N.}~\bibnamefont
  {Spagnolo}}, \bibinfo {author} {\bibfnamefont {C.}~\bibnamefont {Vitelli}},
  \bibinfo {author} {\bibfnamefont {F.}~\bibnamefont {Flamini}}, \bibinfo
  {author} {\bibfnamefont {N.}~\bibnamefont {Viggianiello}}, \bibinfo {author}
  {\bibfnamefont {L.}~\bibnamefont {Latmiral}}, \bibinfo {author}
  {\bibfnamefont {P.}~\bibnamefont {Mataloni}}, \bibinfo {author}
  {\bibfnamefont {D.~J.}\ \bibnamefont {Brod}}, \bibinfo {author}
  {\bibfnamefont {E.~F.}\ \bibnamefont {Galvão}}, \bibinfo {author}
  {\bibfnamefont {A.}~\bibnamefont {Crespi}}, \bibinfo {author} {\bibfnamefont
  {R.}~\bibnamefont {Ramponi}}, \bibinfo {author} {\bibfnamefont
  {R.}~\bibnamefont {Osellame}},\ and\ \bibinfo {author} {\bibfnamefont
  {F.}~\bibnamefont {Sciarrino}},\ }\bibfield  {title} {\bibinfo {title}
  {Experimental scattershot boson sampling},\ }\href
  {https://doi.org/10.1126/sciadv.1400255} {\bibfield  {journal} {\bibinfo
  {journal} {Science Advances}\ }\textbf {\bibinfo {volume} {1}},\ \bibinfo
  {pages} {e1400255} (\bibinfo {year} {2015})}\BibitemShut {NoStop}%
\bibitem [{\citenamefont {Eisaman}\ \emph {et~al.}(2011)\citenamefont
  {Eisaman}, \citenamefont {Fan}, \citenamefont {Migdall},\ and\ \citenamefont
  {Polyakov}}]{eisaman_invited_2011}%
  \BibitemOpen
  \bibfield  {author} {\bibinfo {author} {\bibfnamefont {M.~D.}\ \bibnamefont
  {Eisaman}}, \bibinfo {author} {\bibfnamefont {J.}~\bibnamefont {Fan}},
  \bibinfo {author} {\bibfnamefont {A.}~\bibnamefont {Migdall}},\ and\ \bibinfo
  {author} {\bibfnamefont {S.~V.}\ \bibnamefont {Polyakov}},\ }\bibfield
  {title} {\bibinfo {title} {Invited {Review} {Article}: {Single}-photon
  sources and detectors},\ }\href {https://doi.org/10.1063/1.3610677}
  {\bibfield  {journal} {\bibinfo  {journal} {Review of Scientific
  Instruments}\ }\textbf {\bibinfo {volume} {82}},\ \bibinfo {pages} {071101}
  (\bibinfo {year} {2011})}\BibitemShut {NoStop}%
\bibitem [{\citenamefont {Marsili}\ \emph {et~al.}(2013)\citenamefont
  {Marsili}, \citenamefont {Verma}, \citenamefont {Stern}, \citenamefont
  {Harrington}, \citenamefont {Lita}, \citenamefont {Gerrits}, \citenamefont
  {Vayshenker}, \citenamefont {Baek}, \citenamefont {Shaw}, \citenamefont
  {Mirin},\ and\ \citenamefont {Nam}}]{marsili_detecting_2013}%
  \BibitemOpen
  \bibfield  {author} {\bibinfo {author} {\bibfnamefont {F.}~\bibnamefont
  {Marsili}}, \bibinfo {author} {\bibfnamefont {V.~B.}\ \bibnamefont {Verma}},
  \bibinfo {author} {\bibfnamefont {J.~A.}\ \bibnamefont {Stern}}, \bibinfo
  {author} {\bibfnamefont {S.}~\bibnamefont {Harrington}}, \bibinfo {author}
  {\bibfnamefont {A.~E.}\ \bibnamefont {Lita}}, \bibinfo {author}
  {\bibfnamefont {T.}~\bibnamefont {Gerrits}}, \bibinfo {author} {\bibfnamefont
  {I.}~\bibnamefont {Vayshenker}}, \bibinfo {author} {\bibfnamefont
  {B.}~\bibnamefont {Baek}}, \bibinfo {author} {\bibfnamefont {M.~D.}\
  \bibnamefont {Shaw}}, \bibinfo {author} {\bibfnamefont {R.~P.}\ \bibnamefont
  {Mirin}},\ and\ \bibinfo {author} {\bibfnamefont {S.~W.}\ \bibnamefont
  {Nam}},\ }\bibfield  {title} {\bibinfo {title} {Detecting single infrared
  photons with 93\% system efficiency},\ }\bibfield  {journal} {\bibinfo
  {journal} {Nature Photonics}\ }\textbf {\bibinfo {volume} {7}},\ \href
  {https://doi.org/10.1038/nphoton.2013.13} {10.1038/nphoton.2013.13} (\bibinfo
  {year} {2013})\BibitemShut {NoStop}%
\bibitem [{\citenamefont {Esmaeil~Zadeh}\ \emph {et~al.}(2017)\citenamefont
  {Esmaeil~Zadeh}, \citenamefont {Los}, \citenamefont {Gourgues}, \citenamefont
  {Steinmetz}, \citenamefont {Bulgarini}, \citenamefont {Dobrovolskiy},
  \citenamefont {Zwiller},\ and\ \citenamefont
  {Dorenbos}}]{esmaeil_zadeh_single-photon_2017}%
  \BibitemOpen
  \bibfield  {author} {\bibinfo {author} {\bibfnamefont {I.}~\bibnamefont
  {Esmaeil~Zadeh}}, \bibinfo {author} {\bibfnamefont {J.~W.~N.}\ \bibnamefont
  {Los}}, \bibinfo {author} {\bibfnamefont {R.~B.~M.}\ \bibnamefont
  {Gourgues}}, \bibinfo {author} {\bibfnamefont {V.}~\bibnamefont {Steinmetz}},
  \bibinfo {author} {\bibfnamefont {G.}~\bibnamefont {Bulgarini}}, \bibinfo
  {author} {\bibfnamefont {S.~M.}\ \bibnamefont {Dobrovolskiy}}, \bibinfo
  {author} {\bibfnamefont {V.}~\bibnamefont {Zwiller}},\ and\ \bibinfo {author}
  {\bibfnamefont {S.~N.}\ \bibnamefont {Dorenbos}},\ }\bibfield  {title}
  {\bibinfo {title} {Single-photon detectors combining high efficiency, high
  detection rates, and ultra-high timing resolution},\ }\href
  {https://doi.org/10.1063/1.5000001} {\bibfield  {journal} {\bibinfo
  {journal} {APL Photonics}\ }\textbf {\bibinfo {volume} {2}},\ \bibinfo
  {pages} {111301} (\bibinfo {year} {2017})}\BibitemShut {NoStop}%
\bibitem [{\citenamefont {Simon}\ \emph {et~al.}(2007)\citenamefont {Simon},
  \citenamefont {de~Riedmatten}, \citenamefont {Afzelius}, \citenamefont
  {Sangouard}, \citenamefont {Zbinden},\ and\ \citenamefont
  {Gisin}}]{simon_quantum_2007}%
  \BibitemOpen
  \bibfield  {author} {\bibinfo {author} {\bibfnamefont {C.}~\bibnamefont
  {Simon}}, \bibinfo {author} {\bibfnamefont {H.}~\bibnamefont
  {de~Riedmatten}}, \bibinfo {author} {\bibfnamefont {M.}~\bibnamefont
  {Afzelius}}, \bibinfo {author} {\bibfnamefont {N.}~\bibnamefont {Sangouard}},
  \bibinfo {author} {\bibfnamefont {H.}~\bibnamefont {Zbinden}},\ and\ \bibinfo
  {author} {\bibfnamefont {N.}~\bibnamefont {Gisin}},\ }\bibfield  {title}
  {\bibinfo {title} {Quantum {Repeaters} with {Photon} {Pair} {Sources} and
  {Multimode} {Memories}},\ }\href
  {https://doi.org/10.1103/PhysRevLett.98.190503} {\bibfield  {journal}
  {\bibinfo  {journal} {Phys. Rev. Lett.}\ }\textbf {\bibinfo {volume} {98}},\
  \bibinfo {pages} {190503} (\bibinfo {year} {2007})}\BibitemShut {NoStop}%
\bibitem [{\citenamefont {Cattaneo}\ \emph {et~al.}(2018)\citenamefont
  {Cattaneo}, \citenamefont {Paris},\ and\ \citenamefont
  {Olivares}}]{cattaneo_hybrid_2018}%
  \BibitemOpen
  \bibfield  {author} {\bibinfo {author} {\bibfnamefont {M.}~\bibnamefont
  {Cattaneo}}, \bibinfo {author} {\bibfnamefont {M.~G.~A.}\ \bibnamefont
  {Paris}},\ and\ \bibinfo {author} {\bibfnamefont {S.}~\bibnamefont
  {Olivares}},\ }\bibfield  {title} {\bibinfo {title} {Hybrid quantum key
  distribution using coherent states and photon-number-resolving detectors},\
  }\href {https://doi.org/10.1103/PhysRevA.98.012333} {\bibfield  {journal}
  {\bibinfo  {journal} {Phys. Rev. A}\ }\textbf {\bibinfo {volume} {98}},\
  \bibinfo {pages} {012333} (\bibinfo {year} {2018})}\BibitemShut {NoStop}%
\bibitem [{\citenamefont {Huh}\ \emph {et~al.}(2015)\citenamefont {Huh},
  \citenamefont {Guerreschi}, \citenamefont {Peropadre}, \citenamefont
  {McClean},\ and\ \citenamefont {Aspuru-Guzik}}]{huh_boson_2015}%
  \BibitemOpen
  \bibfield  {author} {\bibinfo {author} {\bibfnamefont {J.}~\bibnamefont
  {Huh}}, \bibinfo {author} {\bibfnamefont {G.~G.}\ \bibnamefont {Guerreschi}},
  \bibinfo {author} {\bibfnamefont {B.}~\bibnamefont {Peropadre}}, \bibinfo
  {author} {\bibfnamefont {J.~R.}\ \bibnamefont {McClean}},\ and\ \bibinfo
  {author} {\bibfnamefont {A.}~\bibnamefont {Aspuru-Guzik}},\ }\bibfield
  {title} {\bibinfo {title} {Boson sampling for molecular vibronic spectra},\
  }\bibfield  {journal} {\bibinfo  {journal} {Nature Photonics}\ }\textbf
  {\bibinfo {volume} {9}},\ \href {https://doi.org/10.1038/nphoton.2015.153}
  {10.1038/nphoton.2015.153} (\bibinfo {year} {2015})\BibitemShut {NoStop}%
\bibitem [{\citenamefont {Hamilton}\ \emph {et~al.}(2017)\citenamefont
  {Hamilton}, \citenamefont {Kruse}, \citenamefont {Sansoni}, \citenamefont
  {Barkhofen}, \citenamefont {Silberhorn},\ and\ \citenamefont
  {Jex}}]{hamilton_gaussian_2017}%
  \BibitemOpen
  \bibfield  {author} {\bibinfo {author} {\bibfnamefont {C.~S.}\ \bibnamefont
  {Hamilton}}, \bibinfo {author} {\bibfnamefont {R.}~\bibnamefont {Kruse}},
  \bibinfo {author} {\bibfnamefont {L.}~\bibnamefont {Sansoni}}, \bibinfo
  {author} {\bibfnamefont {S.}~\bibnamefont {Barkhofen}}, \bibinfo {author}
  {\bibfnamefont {C.}~\bibnamefont {Silberhorn}},\ and\ \bibinfo {author}
  {\bibfnamefont {I.}~\bibnamefont {Jex}},\ }\bibfield  {title} {\bibinfo
  {title} {Gaussian {Boson} {Sampling}},\ }\href
  {https://doi.org/10.1103/PhysRevLett.119.170501} {\bibfield  {journal}
  {\bibinfo  {journal} {Phys. Rev. Lett.}\ }\textbf {\bibinfo {volume} {119}},\
  \bibinfo {pages} {170501} (\bibinfo {year} {2017})}\BibitemShut {NoStop}%
\bibitem [{\citenamefont {Clements}\ \emph {et~al.}(2018)\citenamefont
  {Clements}, \citenamefont {Renema}, \citenamefont {Eckstein}, \citenamefont
  {Valido}, \citenamefont {Lita}, \citenamefont {Gerrits}, \citenamefont {Nam},
  \citenamefont {Kolthammer}, \citenamefont {Huh},\ and\ \citenamefont
  {Walmsley}}]{clements_approximating_2018}%
  \BibitemOpen
  \bibfield  {author} {\bibinfo {author} {\bibfnamefont {W.~R.}\ \bibnamefont
  {Clements}}, \bibinfo {author} {\bibfnamefont {J.~J.}\ \bibnamefont
  {Renema}}, \bibinfo {author} {\bibfnamefont {A.}~\bibnamefont {Eckstein}},
  \bibinfo {author} {\bibfnamefont {A.~A.}\ \bibnamefont {Valido}}, \bibinfo
  {author} {\bibfnamefont {A.}~\bibnamefont {Lita}}, \bibinfo {author}
  {\bibfnamefont {T.}~\bibnamefont {Gerrits}}, \bibinfo {author} {\bibfnamefont
  {S.~W.}\ \bibnamefont {Nam}}, \bibinfo {author} {\bibfnamefont {W.~S.}\
  \bibnamefont {Kolthammer}}, \bibinfo {author} {\bibfnamefont
  {J.}~\bibnamefont {Huh}},\ and\ \bibinfo {author} {\bibfnamefont {I.~A.}\
  \bibnamefont {Walmsley}},\ }\bibfield  {title} {\bibinfo {title}
  {Approximating vibronic spectroscopy with imperfect quantum optics},\ }\href
  {https://doi.org/10.1088/1361-6455/aaf031} {\bibfield  {journal} {\bibinfo
  {journal} {J. Phys. B: At. Mol. Opt. Phys.}\ }\textbf {\bibinfo {volume}
  {51}},\ \bibinfo {pages} {245503} (\bibinfo {year} {2018})}\BibitemShut
  {NoStop}%
\bibitem [{\citenamefont {Kruse}\ \emph {et~al.}(2019)\citenamefont {Kruse},
  \citenamefont {Hamilton}, \citenamefont {Sansoni}, \citenamefont {Barkhofen},
  \citenamefont {Silberhorn},\ and\ \citenamefont {Jex}}]{kruse_detailed_2019}%
  \BibitemOpen
  \bibfield  {author} {\bibinfo {author} {\bibfnamefont {R.}~\bibnamefont
  {Kruse}}, \bibinfo {author} {\bibfnamefont {C.~S.}\ \bibnamefont {Hamilton}},
  \bibinfo {author} {\bibfnamefont {L.}~\bibnamefont {Sansoni}}, \bibinfo
  {author} {\bibfnamefont {S.}~\bibnamefont {Barkhofen}}, \bibinfo {author}
  {\bibfnamefont {C.}~\bibnamefont {Silberhorn}},\ and\ \bibinfo {author}
  {\bibfnamefont {I.}~\bibnamefont {Jex}},\ }\bibfield  {title} {\bibinfo
  {title} {Detailed study of {Gaussian} boson sampling},\ }\href
  {https://doi.org/10.1103/PhysRevA.100.032326} {\bibfield  {journal} {\bibinfo
   {journal} {Phys. Rev. A}\ }\textbf {\bibinfo {volume} {100}},\ \bibinfo
  {pages} {032326} (\bibinfo {year} {2019})}\BibitemShut {NoStop}%
\bibitem [{\citenamefont {Banaszek}\ and\ \citenamefont
  {Walmsley}(2003)}]{banaszek_photon_2003}%
  \BibitemOpen
  \bibfield  {author} {\bibinfo {author} {\bibfnamefont {K.}~\bibnamefont
  {Banaszek}}\ and\ \bibinfo {author} {\bibfnamefont {I.~A.}\ \bibnamefont
  {Walmsley}},\ }\bibfield  {title} {\bibinfo {title} {Photon counting with a
  loop detector},\ }\href {https://doi.org/10.1364/OL.28.000052} {\bibfield
  {journal} {\bibinfo  {journal} {Opt. Lett., OL}\ }\textbf {\bibinfo {volume}
  {28}},\ \bibinfo {pages} {52} (\bibinfo {year} {2003})}\BibitemShut {NoStop}%
\bibitem [{\citenamefont {Fitch}\ \emph {et~al.}(2003)\citenamefont {Fitch},
  \citenamefont {Jacobs}, \citenamefont {Pittman},\ and\ \citenamefont
  {Franson}}]{fitch_photon-number_2003}%
  \BibitemOpen
  \bibfield  {author} {\bibinfo {author} {\bibfnamefont {M.~J.}\ \bibnamefont
  {Fitch}}, \bibinfo {author} {\bibfnamefont {B.~C.}\ \bibnamefont {Jacobs}},
  \bibinfo {author} {\bibfnamefont {T.~B.}\ \bibnamefont {Pittman}},\ and\
  \bibinfo {author} {\bibfnamefont {J.~D.}\ \bibnamefont {Franson}},\
  }\bibfield  {title} {\bibinfo {title} {Photon-number resolution using
  time-multiplexed single-photon detectors},\ }\href
  {https://doi.org/10.1103/PhysRevA.68.043814} {\bibfield  {journal} {\bibinfo
  {journal} {Phys. Rev. A}\ }\textbf {\bibinfo {volume} {68}},\ \bibinfo
  {pages} {043814} (\bibinfo {year} {2003})}\BibitemShut {NoStop}%
\bibitem [{\citenamefont {Achilles}\ \emph {et~al.}(2004)\citenamefont
  {Achilles}, \citenamefont {Silberhorn}, \citenamefont {Sliwa}, \citenamefont
  {Banaszek}, \citenamefont {Walmsley}, \citenamefont {Fitch}, \citenamefont
  {Jacobs}, \citenamefont {Pittman},\ and\ \citenamefont
  {Franson}}]{achilles_photon-number-resolving_2004}%
  \BibitemOpen
  \bibfield  {author} {\bibinfo {author} {\bibfnamefont {D.}~\bibnamefont
  {Achilles}}, \bibinfo {author} {\bibfnamefont {C.}~\bibnamefont
  {Silberhorn}}, \bibinfo {author} {\bibfnamefont {C.}~\bibnamefont {Sliwa}},
  \bibinfo {author} {\bibfnamefont {K.}~\bibnamefont {Banaszek}}, \bibinfo
  {author} {\bibfnamefont {I.~A.}\ \bibnamefont {Walmsley}}, \bibinfo {author}
  {\bibfnamefont {M.~J.}\ \bibnamefont {Fitch}}, \bibinfo {author}
  {\bibfnamefont {B.~C.}\ \bibnamefont {Jacobs}}, \bibinfo {author}
  {\bibfnamefont {T.~B.}\ \bibnamefont {Pittman}},\ and\ \bibinfo {author}
  {\bibfnamefont {J.~D.}\ \bibnamefont {Franson}},\ }\bibfield  {title}
  {\bibinfo {title} {Photon-number-resolving detection using
  time-multiplexing},\ }\href {https://doi.org/10.1080/09500340408235288}
  {\bibfield  {journal} {\bibinfo  {journal} {Journal of Modern Optics}\
  }\textbf {\bibinfo {volume} {51}},\ \bibinfo {pages} {1499} (\bibinfo {year}
  {2004})}\BibitemShut {NoStop}%
\bibitem [{\citenamefont {Divochiy}\ \emph {et~al.}(2008)\citenamefont
  {Divochiy}, \citenamefont {Marsili}, \citenamefont {Bitauld}, \citenamefont
  {Gaggero}, \citenamefont {Leoni}, \citenamefont {Mattioli}, \citenamefont
  {Korneev}, \citenamefont {Seleznev}, \citenamefont {Kaurova}, \citenamefont
  {Minaeva}, \citenamefont {Gol'tsman}, \citenamefont {Lagoudakis},
  \citenamefont {Benkhaoul}, \citenamefont {Lévy},\ and\ \citenamefont
  {Fiore}}]{divochiy_superconducting_2008}%
  \BibitemOpen
  \bibfield  {author} {\bibinfo {author} {\bibfnamefont {A.}~\bibnamefont
  {Divochiy}}, \bibinfo {author} {\bibfnamefont {F.}~\bibnamefont {Marsili}},
  \bibinfo {author} {\bibfnamefont {D.}~\bibnamefont {Bitauld}}, \bibinfo
  {author} {\bibfnamefont {A.}~\bibnamefont {Gaggero}}, \bibinfo {author}
  {\bibfnamefont {R.}~\bibnamefont {Leoni}}, \bibinfo {author} {\bibfnamefont
  {F.}~\bibnamefont {Mattioli}}, \bibinfo {author} {\bibfnamefont
  {A.}~\bibnamefont {Korneev}}, \bibinfo {author} {\bibfnamefont
  {V.}~\bibnamefont {Seleznev}}, \bibinfo {author} {\bibfnamefont
  {N.}~\bibnamefont {Kaurova}}, \bibinfo {author} {\bibfnamefont
  {O.}~\bibnamefont {Minaeva}}, \bibinfo {author} {\bibfnamefont
  {G.}~\bibnamefont {Gol'tsman}}, \bibinfo {author} {\bibfnamefont {K.~G.}\
  \bibnamefont {Lagoudakis}}, \bibinfo {author} {\bibfnamefont
  {M.}~\bibnamefont {Benkhaoul}}, \bibinfo {author} {\bibfnamefont
  {F.}~\bibnamefont {Lévy}},\ and\ \bibinfo {author} {\bibfnamefont
  {A.}~\bibnamefont {Fiore}},\ }\bibfield  {title} {\bibinfo {title}
  {Superconducting nanowire photon-number-resolving detector at
  telecommunication wavelengths},\ }\bibfield  {journal} {\bibinfo  {journal}
  {Nature Photonics}\ }\textbf {\bibinfo {volume} {2}},\ \href
  {https://doi.org/10.1038/nphoton.2008.51} {10.1038/nphoton.2008.51} (\bibinfo
  {year} {2008})\BibitemShut {NoStop}%
\bibitem [{\citenamefont {Mattioli}\ \emph {et~al.}(2016)\citenamefont
  {Mattioli}, \citenamefont {Zhou}, \citenamefont {Gaggero}, \citenamefont
  {Gaudio}, \citenamefont {Leoni},\ and\ \citenamefont
  {Fiore}}]{mattioli_photon-counting_2016}%
  \BibitemOpen
  \bibfield  {author} {\bibinfo {author} {\bibfnamefont {F.}~\bibnamefont
  {Mattioli}}, \bibinfo {author} {\bibfnamefont {Z.}~\bibnamefont {Zhou}},
  \bibinfo {author} {\bibfnamefont {A.}~\bibnamefont {Gaggero}}, \bibinfo
  {author} {\bibfnamefont {R.}~\bibnamefont {Gaudio}}, \bibinfo {author}
  {\bibfnamefont {R.}~\bibnamefont {Leoni}},\ and\ \bibinfo {author}
  {\bibfnamefont {A.}~\bibnamefont {Fiore}},\ }\bibfield  {title} {\bibinfo
  {title} {Photon-counting and analog operation of a 24-pixel photon number
  resolving detector based on superconducting nanowires},\ }\href
  {https://doi.org/10.1364/OE.24.009067} {\bibfield  {journal} {\bibinfo
  {journal} {Opt. Express, OE}\ }\textbf {\bibinfo {volume} {24}},\ \bibinfo
  {pages} {9067} (\bibinfo {year} {2016})}\BibitemShut {NoStop}%
\bibitem [{\citenamefont {Tao}\ \emph {et~al.}(2019)\citenamefont {Tao},
  \citenamefont {Chen}, \citenamefont {Chen}, \citenamefont {Wang},
  \citenamefont {Li}, \citenamefont {Tu}, \citenamefont {Jia}, \citenamefont
  {Zhao}, \citenamefont {Zhang}, \citenamefont {Kang},\ and\ \citenamefont
  {Wu}}]{tao_high_2019}%
  \BibitemOpen
  \bibfield  {author} {\bibinfo {author} {\bibfnamefont {X.}~\bibnamefont
  {Tao}}, \bibinfo {author} {\bibfnamefont {S.}~\bibnamefont {Chen}}, \bibinfo
  {author} {\bibfnamefont {Y.}~\bibnamefont {Chen}}, \bibinfo {author}
  {\bibfnamefont {L.}~\bibnamefont {Wang}}, \bibinfo {author} {\bibfnamefont
  {X.}~\bibnamefont {Li}}, \bibinfo {author} {\bibfnamefont {X.}~\bibnamefont
  {Tu}}, \bibinfo {author} {\bibfnamefont {X.}~\bibnamefont {Jia}}, \bibinfo
  {author} {\bibfnamefont {Q.}~\bibnamefont {Zhao}}, \bibinfo {author}
  {\bibfnamefont {L.}~\bibnamefont {Zhang}}, \bibinfo {author} {\bibfnamefont
  {L.}~\bibnamefont {Kang}},\ and\ \bibinfo {author} {\bibfnamefont
  {P.}~\bibnamefont {Wu}},\ }\bibfield  {title} {\bibinfo {title} {A high speed
  and high efficiency superconducting photon number resolving detector},\
  }\href {https://doi.org/10.1088/1361-6668/ab0799} {\bibfield  {journal}
  {\bibinfo  {journal} {Supercond. Sci. Technol.}\ }\textbf {\bibinfo {volume}
  {32}},\ \bibinfo {pages} {064002} (\bibinfo {year} {2019})}\BibitemShut
  {NoStop}%
\bibitem [{\citenamefont {Kardynał}\ \emph {et~al.}(2008)\citenamefont
  {Kardynał}, \citenamefont {Yuan},\ and\ \citenamefont
  {Shields}}]{kardynal_avalanchephotodiode-based_2008}%
  \BibitemOpen
  \bibfield  {author} {\bibinfo {author} {\bibfnamefont {B.~E.}\ \bibnamefont
  {Kardynał}}, \bibinfo {author} {\bibfnamefont {Z.~L.}\ \bibnamefont
  {Yuan}},\ and\ \bibinfo {author} {\bibfnamefont {A.~J.}\ \bibnamefont
  {Shields}},\ }\bibfield  {title} {\bibinfo {title} {An
  avalanche‐photodiode-based photon-number-resolving detector},\ }\bibfield
  {journal} {\bibinfo  {journal} {Nature Photonics}\ }\textbf {\bibinfo
  {volume} {2}},\ \href {https://doi.org/10.1038/nphoton.2008.101}
  {10.1038/nphoton.2008.101} (\bibinfo {year} {2008})\BibitemShut {NoStop}%
\bibitem [{\citenamefont {Gerrits}\ \emph {et~al.}(2011)\citenamefont
  {Gerrits}, \citenamefont {Thomas-Peter}, \citenamefont {Gates}, \citenamefont
  {Lita}, \citenamefont {Metcalf}, \citenamefont {Calkins}, \citenamefont
  {Tomlin}, \citenamefont {Fox}, \citenamefont {Linares}, \citenamefont
  {Spring}, \citenamefont {Langford}, \citenamefont {Mirin}, \citenamefont
  {Smith}, \citenamefont {Walmsley},\ and\ \citenamefont
  {Nam}}]{gerrits_-chip_2011}%
  \BibitemOpen
  \bibfield  {author} {\bibinfo {author} {\bibfnamefont {T.}~\bibnamefont
  {Gerrits}}, \bibinfo {author} {\bibfnamefont {N.}~\bibnamefont
  {Thomas-Peter}}, \bibinfo {author} {\bibfnamefont {J.~C.}\ \bibnamefont
  {Gates}}, \bibinfo {author} {\bibfnamefont {A.~E.}\ \bibnamefont {Lita}},
  \bibinfo {author} {\bibfnamefont {B.~J.}\ \bibnamefont {Metcalf}}, \bibinfo
  {author} {\bibfnamefont {B.}~\bibnamefont {Calkins}}, \bibinfo {author}
  {\bibfnamefont {N.~A.}\ \bibnamefont {Tomlin}}, \bibinfo {author}
  {\bibfnamefont {A.~E.}\ \bibnamefont {Fox}}, \bibinfo {author} {\bibfnamefont
  {A.~L.}\ \bibnamefont {Linares}}, \bibinfo {author} {\bibfnamefont {J.~B.}\
  \bibnamefont {Spring}}, \bibinfo {author} {\bibfnamefont {N.~K.}\
  \bibnamefont {Langford}}, \bibinfo {author} {\bibfnamefont {R.~P.}\
  \bibnamefont {Mirin}}, \bibinfo {author} {\bibfnamefont {P.~G.~R.}\
  \bibnamefont {Smith}}, \bibinfo {author} {\bibfnamefont {I.~A.}\ \bibnamefont
  {Walmsley}},\ and\ \bibinfo {author} {\bibfnamefont {S.~W.}\ \bibnamefont
  {Nam}},\ }\bibfield  {title} {\bibinfo {title} {On-chip,
  photon-number-resolving, telecommunication-band detectors for scalable
  photonic information processing},\ }\href
  {https://doi.org/10.1103/PhysRevA.84.060301} {\bibfield  {journal} {\bibinfo
  {journal} {Phys. Rev. A}\ }\textbf {\bibinfo {volume} {84}},\ \bibinfo
  {pages} {060301(R)} (\bibinfo {year} {2011})}\BibitemShut {NoStop}%
\bibitem [{\citenamefont {Calkins}\ \emph {et~al.}(2013)\citenamefont
  {Calkins}, \citenamefont {Mennea}, \citenamefont {Lita}, \citenamefont
  {Metcalf}, \citenamefont {Kolthammer}, \citenamefont {Linares}, \citenamefont
  {Spring}, \citenamefont {Humphreys}, \citenamefont {Mirin}, \citenamefont
  {Gates}, \citenamefont {Smith}, \citenamefont {Walmsley}, \citenamefont
  {Gerrits},\ and\ \citenamefont {Nam}}]{calkins_high_2013}%
  \BibitemOpen
  \bibfield  {author} {\bibinfo {author} {\bibfnamefont {B.}~\bibnamefont
  {Calkins}}, \bibinfo {author} {\bibfnamefont {P.~L.}\ \bibnamefont {Mennea}},
  \bibinfo {author} {\bibfnamefont {A.~E.}\ \bibnamefont {Lita}}, \bibinfo
  {author} {\bibfnamefont {B.~J.}\ \bibnamefont {Metcalf}}, \bibinfo {author}
  {\bibfnamefont {W.~S.}\ \bibnamefont {Kolthammer}}, \bibinfo {author}
  {\bibfnamefont {A.~L.}\ \bibnamefont {Linares}}, \bibinfo {author}
  {\bibfnamefont {J.~B.}\ \bibnamefont {Spring}}, \bibinfo {author}
  {\bibfnamefont {P.~C.}\ \bibnamefont {Humphreys}}, \bibinfo {author}
  {\bibfnamefont {R.~P.}\ \bibnamefont {Mirin}}, \bibinfo {author}
  {\bibfnamefont {J.~C.}\ \bibnamefont {Gates}}, \bibinfo {author}
  {\bibfnamefont {P.~G.~R.}\ \bibnamefont {Smith}}, \bibinfo {author}
  {\bibfnamefont {I.~A.}\ \bibnamefont {Walmsley}}, \bibinfo {author}
  {\bibfnamefont {T.}~\bibnamefont {Gerrits}},\ and\ \bibinfo {author}
  {\bibfnamefont {S.~W.}\ \bibnamefont {Nam}},\ }\bibfield  {title} {\bibinfo
  {title} {High quantum-efficiency photon-number-resolving detector for
  photonic on-chip information processing},\ }\href
  {https://doi.org/10.1364/OE.21.022657} {\bibfield  {journal} {\bibinfo
  {journal} {Opt. Express}\ }\textbf {\bibinfo {volume} {21}},\ \bibinfo
  {pages} {22657} (\bibinfo {year} {2013})}\BibitemShut {NoStop}%
\bibitem [{\citenamefont {Gleyzes}\ \emph {et~al.}(2007)\citenamefont
  {Gleyzes}, \citenamefont {Kuhr}, \citenamefont {Guerlin}, \citenamefont
  {Bernu}, \citenamefont {Deléglise}, \citenamefont {Busk~Hoff}, \citenamefont
  {Brune}, \citenamefont {Raimond},\ and\ \citenamefont
  {Haroche}}]{gleyzes_quantum_2007}%
  \BibitemOpen
  \bibfield  {author} {\bibinfo {author} {\bibfnamefont {S.}~\bibnamefont
  {Gleyzes}}, \bibinfo {author} {\bibfnamefont {S.}~\bibnamefont {Kuhr}},
  \bibinfo {author} {\bibfnamefont {C.}~\bibnamefont {Guerlin}}, \bibinfo
  {author} {\bibfnamefont {J.}~\bibnamefont {Bernu}}, \bibinfo {author}
  {\bibfnamefont {S.}~\bibnamefont {Deléglise}}, \bibinfo {author}
  {\bibfnamefont {U.}~\bibnamefont {Busk~Hoff}}, \bibinfo {author}
  {\bibfnamefont {M.}~\bibnamefont {Brune}}, \bibinfo {author} {\bibfnamefont
  {J.-M.}\ \bibnamefont {Raimond}},\ and\ \bibinfo {author} {\bibfnamefont
  {S.}~\bibnamefont {Haroche}},\ }\bibfield  {title} {\bibinfo {title} {Quantum
  jumps of light recording the birth and death of a photon in a cavity},\
  }\bibfield  {journal} {\bibinfo  {journal} {Nature}\ }\textbf {\bibinfo
  {volume} {446}},\ \href {https://doi.org/10.1038/nature05589}
  {10.1038/nature05589} (\bibinfo {year} {2007})\BibitemShut {NoStop}%
\bibitem [{\citenamefont {Schuster}\ \emph {et~al.}(2007)\citenamefont
  {Schuster}, \citenamefont {Houck}, \citenamefont {Schreier}, \citenamefont
  {Wallraff}, \citenamefont {Gambetta}, \citenamefont {Blais}, \citenamefont
  {Frunzio}, \citenamefont {Majer}, \citenamefont {Johnson}, \citenamefont
  {Devoret}, \citenamefont {Girvin},\ and\ \citenamefont
  {Schoelkopf}}]{schuster_resolving_2007}%
  \BibitemOpen
  \bibfield  {author} {\bibinfo {author} {\bibfnamefont {D.~I.}\ \bibnamefont
  {Schuster}}, \bibinfo {author} {\bibfnamefont {A.~A.}\ \bibnamefont {Houck}},
  \bibinfo {author} {\bibfnamefont {J.~A.}\ \bibnamefont {Schreier}}, \bibinfo
  {author} {\bibfnamefont {A.}~\bibnamefont {Wallraff}}, \bibinfo {author}
  {\bibfnamefont {J.~M.}\ \bibnamefont {Gambetta}}, \bibinfo {author}
  {\bibfnamefont {A.}~\bibnamefont {Blais}}, \bibinfo {author} {\bibfnamefont
  {L.}~\bibnamefont {Frunzio}}, \bibinfo {author} {\bibfnamefont
  {J.}~\bibnamefont {Majer}}, \bibinfo {author} {\bibfnamefont
  {B.}~\bibnamefont {Johnson}}, \bibinfo {author} {\bibfnamefont {M.~H.}\
  \bibnamefont {Devoret}}, \bibinfo {author} {\bibfnamefont {S.~M.}\
  \bibnamefont {Girvin}},\ and\ \bibinfo {author} {\bibfnamefont {R.~J.}\
  \bibnamefont {Schoelkopf}},\ }\bibfield  {title} {\bibinfo {title} {Resolving
  photon number states in a superconducting circuit},\ }\bibfield  {journal}
  {\bibinfo  {journal} {Nature}\ }\textbf {\bibinfo {volume} {445}},\ \href
  {https://doi.org/10.1038/nature05461} {10.1038/nature05461} (\bibinfo {year}
  {2007})\BibitemShut {NoStop}%
\bibitem [{\citenamefont {Johnson}\ \emph {et~al.}(2010)\citenamefont
  {Johnson}, \citenamefont {Reed}, \citenamefont {Houck}, \citenamefont
  {Schuster}, \citenamefont {Bishop}, \citenamefont {Ginossar}, \citenamefont
  {Gambetta}, \citenamefont {DiCarlo}, \citenamefont {Frunzio}, \citenamefont
  {Girvin},\ and\ \citenamefont {Schoelkopf}}]{johnson_quantum_2010}%
  \BibitemOpen
  \bibfield  {author} {\bibinfo {author} {\bibfnamefont {B.~R.}\ \bibnamefont
  {Johnson}}, \bibinfo {author} {\bibfnamefont {M.~D.}\ \bibnamefont {Reed}},
  \bibinfo {author} {\bibfnamefont {A.~A.}\ \bibnamefont {Houck}}, \bibinfo
  {author} {\bibfnamefont {D.~I.}\ \bibnamefont {Schuster}}, \bibinfo {author}
  {\bibfnamefont {L.~S.}\ \bibnamefont {Bishop}}, \bibinfo {author}
  {\bibfnamefont {E.}~\bibnamefont {Ginossar}}, \bibinfo {author}
  {\bibfnamefont {J.~M.}\ \bibnamefont {Gambetta}}, \bibinfo {author}
  {\bibfnamefont {L.}~\bibnamefont {DiCarlo}}, \bibinfo {author} {\bibfnamefont
  {L.}~\bibnamefont {Frunzio}}, \bibinfo {author} {\bibfnamefont {S.~M.}\
  \bibnamefont {Girvin}},\ and\ \bibinfo {author} {\bibfnamefont {R.~J.}\
  \bibnamefont {Schoelkopf}},\ }\bibfield  {title} {\bibinfo {title} {Quantum
  non-demolition detection of single microwave photons in a circuit},\
  }\bibfield  {journal} {\bibinfo  {journal} {Nature Physics}\ }\textbf
  {\bibinfo {volume} {6}},\ \href {https://doi.org/10.1038/nphys1710}
  {10.1038/nphys1710} (\bibinfo {year} {2010})\BibitemShut {NoStop}%
\bibitem [{\citenamefont {Houck}\ \emph {et~al.}(2007)\citenamefont {Houck},
  \citenamefont {Schuster}, \citenamefont {Gambetta}, \citenamefont {Schreier},
  \citenamefont {Johnson}, \citenamefont {Chow}, \citenamefont {Frunzio},
  \citenamefont {Majer}, \citenamefont {Devoret}, \citenamefont {Girvin},\ and\
  \citenamefont {Schoelkopf}}]{houck_generating_2007}%
  \BibitemOpen
  \bibfield  {author} {\bibinfo {author} {\bibfnamefont {A.~A.}\ \bibnamefont
  {Houck}}, \bibinfo {author} {\bibfnamefont {D.~I.}\ \bibnamefont {Schuster}},
  \bibinfo {author} {\bibfnamefont {J.~M.}\ \bibnamefont {Gambetta}}, \bibinfo
  {author} {\bibfnamefont {J.~A.}\ \bibnamefont {Schreier}}, \bibinfo {author}
  {\bibfnamefont {B.~R.}\ \bibnamefont {Johnson}}, \bibinfo {author}
  {\bibfnamefont {J.~M.}\ \bibnamefont {Chow}}, \bibinfo {author}
  {\bibfnamefont {L.}~\bibnamefont {Frunzio}}, \bibinfo {author} {\bibfnamefont
  {J.}~\bibnamefont {Majer}}, \bibinfo {author} {\bibfnamefont {M.~H.}\
  \bibnamefont {Devoret}}, \bibinfo {author} {\bibfnamefont {S.~M.}\
  \bibnamefont {Girvin}},\ and\ \bibinfo {author} {\bibfnamefont {R.~J.}\
  \bibnamefont {Schoelkopf}},\ }\bibfield  {title} {\bibinfo {title}
  {Generating single microwave photons in a circuit},\ }\bibfield  {journal}
  {\bibinfo  {journal} {Nature}\ }\textbf {\bibinfo {volume} {449}},\ \href
  {https://doi.org/10.1038/nature06126} {10.1038/nature06126} (\bibinfo {year}
  {2007})\BibitemShut {NoStop}%
\bibitem [{\citenamefont {Hofheinz}\ \emph {et~al.}(2008)\citenamefont
  {Hofheinz}, \citenamefont {Weig}, \citenamefont {Ansmann}, \citenamefont
  {Bialczak}, \citenamefont {Lucero}, \citenamefont {Neeley}, \citenamefont
  {O’Connell}, \citenamefont {Wang}, \citenamefont {Martinis},\ and\
  \citenamefont {Cleland}}]{hofheinz_generation_2008}%
  \BibitemOpen
  \bibfield  {author} {\bibinfo {author} {\bibfnamefont {M.}~\bibnamefont
  {Hofheinz}}, \bibinfo {author} {\bibfnamefont {E.~M.}\ \bibnamefont {Weig}},
  \bibinfo {author} {\bibfnamefont {M.}~\bibnamefont {Ansmann}}, \bibinfo
  {author} {\bibfnamefont {R.~C.}\ \bibnamefont {Bialczak}}, \bibinfo {author}
  {\bibfnamefont {E.}~\bibnamefont {Lucero}}, \bibinfo {author} {\bibfnamefont
  {M.}~\bibnamefont {Neeley}}, \bibinfo {author} {\bibfnamefont {A.~D.}\
  \bibnamefont {O’Connell}}, \bibinfo {author} {\bibfnamefont
  {H.}~\bibnamefont {Wang}}, \bibinfo {author} {\bibfnamefont {J.~M.}\
  \bibnamefont {Martinis}},\ and\ \bibinfo {author} {\bibfnamefont {A.~N.}\
  \bibnamefont {Cleland}},\ }\bibfield  {title} {\bibinfo {title} {Generation
  of {Fock} states in a superconducting quantum circuit},\ }\bibfield
  {journal} {\bibinfo  {journal} {Nature}\ }\textbf {\bibinfo {volume} {454}},\
  \href {https://doi.org/10.1038/nature07136} {10.1038/nature07136} (\bibinfo
  {year} {2008})\BibitemShut {NoStop}%
\bibitem [{\citenamefont {Heeres}\ \emph {et~al.}(2015)\citenamefont {Heeres},
  \citenamefont {Vlastakis}, \citenamefont {Holland}, \citenamefont
  {Krastanov}, \citenamefont {Albert}, \citenamefont {Frunzio}, \citenamefont
  {Jiang},\ and\ \citenamefont {Schoelkopf}}]{heeres_cavity_2015}%
  \BibitemOpen
  \bibfield  {author} {\bibinfo {author} {\bibfnamefont {R.~W.}\ \bibnamefont
  {Heeres}}, \bibinfo {author} {\bibfnamefont {B.}~\bibnamefont {Vlastakis}},
  \bibinfo {author} {\bibfnamefont {E.}~\bibnamefont {Holland}}, \bibinfo
  {author} {\bibfnamefont {S.}~\bibnamefont {Krastanov}}, \bibinfo {author}
  {\bibfnamefont {V.~V.}\ \bibnamefont {Albert}}, \bibinfo {author}
  {\bibfnamefont {L.}~\bibnamefont {Frunzio}}, \bibinfo {author} {\bibfnamefont
  {L.}~\bibnamefont {Jiang}},\ and\ \bibinfo {author} {\bibfnamefont {R.~J.}\
  \bibnamefont {Schoelkopf}},\ }\bibfield  {title} {\bibinfo {title} {Cavity
  {State} {Manipulation} {Using} {Photon}-{Number} {Selective} {Phase}
  {Gates}},\ }\href {https://doi.org/10.1103/PhysRevLett.115.137002} {\bibfield
   {journal} {\bibinfo  {journal} {Phys. Rev. Lett.}\ }\textbf {\bibinfo
  {volume} {115}},\ \bibinfo {pages} {137002} (\bibinfo {year}
  {2015})}\BibitemShut {NoStop}%
\bibitem [{\citenamefont {Heeres}\ \emph {et~al.}(2017)\citenamefont {Heeres},
  \citenamefont {Reinhold}, \citenamefont {Ofek}, \citenamefont {Frunzio},
  \citenamefont {Jiang}, \citenamefont {Devoret},\ and\ \citenamefont
  {Schoelkopf}}]{heeres_implementing_2017}%
  \BibitemOpen
  \bibfield  {author} {\bibinfo {author} {\bibfnamefont {R.~W.}\ \bibnamefont
  {Heeres}}, \bibinfo {author} {\bibfnamefont {P.}~\bibnamefont {Reinhold}},
  \bibinfo {author} {\bibfnamefont {N.}~\bibnamefont {Ofek}}, \bibinfo {author}
  {\bibfnamefont {L.}~\bibnamefont {Frunzio}}, \bibinfo {author} {\bibfnamefont
  {L.}~\bibnamefont {Jiang}}, \bibinfo {author} {\bibfnamefont {M.~H.}\
  \bibnamefont {Devoret}},\ and\ \bibinfo {author} {\bibfnamefont {R.~J.}\
  \bibnamefont {Schoelkopf}},\ }\bibfield  {title} {\bibinfo {title}
  {Implementing a universal gate set on a logical qubit encoded in an
  oscillator},\ }\bibfield  {journal} {\bibinfo  {journal} {Nature
  Communications}\ }\textbf {\bibinfo {volume} {8}},\ \href
  {https://doi.org/10.1038/s41467-017-00045-1} {10.1038/s41467-017-00045-1}
  (\bibinfo {year} {2017})\BibitemShut {NoStop}%
\bibitem [{\citenamefont {Chen}\ \emph {et~al.}(2011)\citenamefont {Chen},
  \citenamefont {Hover}, \citenamefont {Sendelbach}, \citenamefont {Maurer},
  \citenamefont {Merkel}, \citenamefont {Pritchett}, \citenamefont {Wilhelm},\
  and\ \citenamefont {McDermott}}]{chen_microwave_2011}%
  \BibitemOpen
  \bibfield  {author} {\bibinfo {author} {\bibfnamefont {Y.-F.}\ \bibnamefont
  {Chen}}, \bibinfo {author} {\bibfnamefont {D.}~\bibnamefont {Hover}},
  \bibinfo {author} {\bibfnamefont {S.}~\bibnamefont {Sendelbach}}, \bibinfo
  {author} {\bibfnamefont {L.}~\bibnamefont {Maurer}}, \bibinfo {author}
  {\bibfnamefont {S.~T.}\ \bibnamefont {Merkel}}, \bibinfo {author}
  {\bibfnamefont {E.~J.}\ \bibnamefont {Pritchett}}, \bibinfo {author}
  {\bibfnamefont {F.~K.}\ \bibnamefont {Wilhelm}},\ and\ \bibinfo {author}
  {\bibfnamefont {R.}~\bibnamefont {McDermott}},\ }\bibfield  {title} {\bibinfo
  {title} {Microwave {Photon} {Counter} {Based} on {Josephson} {Junctions}},\
  }\href {https://doi.org/10.1103/PhysRevLett.107.217401} {\bibfield  {journal}
  {\bibinfo  {journal} {Phys. Rev. Lett.}\ }\textbf {\bibinfo {volume} {107}},\
  \bibinfo {pages} {217401} (\bibinfo {year} {2011})}\BibitemShut {NoStop}%
\bibitem [{\citenamefont {Govia}\ \emph {et~al.}(2014)\citenamefont {Govia},
  \citenamefont {Pritchett}, \citenamefont {Xu}, \citenamefont {Plourde},
  \citenamefont {Vavilov}, \citenamefont {Wilhelm},\ and\ \citenamefont
  {McDermott}}]{govia_high-fidelity_2014}%
  \BibitemOpen
  \bibfield  {author} {\bibinfo {author} {\bibfnamefont {L.~C.~G.}\
  \bibnamefont {Govia}}, \bibinfo {author} {\bibfnamefont {E.~J.}\ \bibnamefont
  {Pritchett}}, \bibinfo {author} {\bibfnamefont {C.}~\bibnamefont {Xu}},
  \bibinfo {author} {\bibfnamefont {B.~L.~T.}\ \bibnamefont {Plourde}},
  \bibinfo {author} {\bibfnamefont {M.~G.}\ \bibnamefont {Vavilov}}, \bibinfo
  {author} {\bibfnamefont {F.~K.}\ \bibnamefont {Wilhelm}},\ and\ \bibinfo
  {author} {\bibfnamefont {R.}~\bibnamefont {McDermott}},\ }\bibfield  {title}
  {\bibinfo {title} {High-fidelity qubit measurement with a microwave-photon
  counter},\ }\href {https://doi.org/10.1103/PhysRevA.90.062307} {\bibfield
  {journal} {\bibinfo  {journal} {Phys. Rev. A}\ }\textbf {\bibinfo {volume}
  {90}},\ \bibinfo {pages} {062307} (\bibinfo {year} {2014})}\BibitemShut
  {NoStop}%
\bibitem [{\citenamefont {Opremcak}\ \emph {et~al.}(2018)\citenamefont
  {Opremcak}, \citenamefont {Pechenezhskiy}, \citenamefont {Howington},
  \citenamefont {Christensen}, \citenamefont {Beck}, \citenamefont {Leonard},
  \citenamefont {Suttle}, \citenamefont {Wilen}, \citenamefont {Nesterov},
  \citenamefont {Ribeill}, \citenamefont {Thorbeck}, \citenamefont {Schlenker},
  \citenamefont {Vavilov}, \citenamefont {Plourde},\ and\ \citenamefont
  {McDermott}}]{opremcak_measurement_2018}%
  \BibitemOpen
  \bibfield  {author} {\bibinfo {author} {\bibfnamefont {A.}~\bibnamefont
  {Opremcak}}, \bibinfo {author} {\bibfnamefont {I.~V.}\ \bibnamefont
  {Pechenezhskiy}}, \bibinfo {author} {\bibfnamefont {C.}~\bibnamefont
  {Howington}}, \bibinfo {author} {\bibfnamefont {B.~G.}\ \bibnamefont
  {Christensen}}, \bibinfo {author} {\bibfnamefont {M.~A.}\ \bibnamefont
  {Beck}}, \bibinfo {author} {\bibfnamefont {E.}~\bibnamefont {Leonard}},
  \bibinfo {author} {\bibfnamefont {J.}~\bibnamefont {Suttle}}, \bibinfo
  {author} {\bibfnamefont {C.}~\bibnamefont {Wilen}}, \bibinfo {author}
  {\bibfnamefont {K.~N.}\ \bibnamefont {Nesterov}}, \bibinfo {author}
  {\bibfnamefont {G.~J.}\ \bibnamefont {Ribeill}}, \bibinfo {author}
  {\bibfnamefont {T.}~\bibnamefont {Thorbeck}}, \bibinfo {author}
  {\bibfnamefont {F.}~\bibnamefont {Schlenker}}, \bibinfo {author}
  {\bibfnamefont {M.~G.}\ \bibnamefont {Vavilov}}, \bibinfo {author}
  {\bibfnamefont {B.~L.~T.}\ \bibnamefont {Plourde}},\ and\ \bibinfo {author}
  {\bibfnamefont {R.}~\bibnamefont {McDermott}},\ }\bibfield  {title} {\bibinfo
  {title} {Measurement of a superconducting qubit with a microwave photon
  counter},\ }\href {https://doi.org/10.1126/science.aat4625} {\bibfield
  {journal} {\bibinfo  {journal} {Science}\ }\textbf {\bibinfo {volume}
  {361}},\ \bibinfo {pages} {1239} (\bibinfo {year} {2018})}\BibitemShut
  {NoStop}%
\bibitem [{\citenamefont {Guerlin}\ \emph {et~al.}(2007)\citenamefont
  {Guerlin}, \citenamefont {Bernu}, \citenamefont {Deléglise}, \citenamefont
  {Sayrin}, \citenamefont {Gleyzes}, \citenamefont {Kuhr}, \citenamefont
  {Brune}, \citenamefont {Raimond},\ and\ \citenamefont
  {Haroche}}]{guerlin_progressive_2007}%
  \BibitemOpen
  \bibfield  {author} {\bibinfo {author} {\bibfnamefont {C.}~\bibnamefont
  {Guerlin}}, \bibinfo {author} {\bibfnamefont {J.}~\bibnamefont {Bernu}},
  \bibinfo {author} {\bibfnamefont {S.}~\bibnamefont {Deléglise}}, \bibinfo
  {author} {\bibfnamefont {C.}~\bibnamefont {Sayrin}}, \bibinfo {author}
  {\bibfnamefont {S.}~\bibnamefont {Gleyzes}}, \bibinfo {author} {\bibfnamefont
  {S.}~\bibnamefont {Kuhr}}, \bibinfo {author} {\bibfnamefont {M.}~\bibnamefont
  {Brune}}, \bibinfo {author} {\bibfnamefont {J.-M.}\ \bibnamefont {Raimond}},\
  and\ \bibinfo {author} {\bibfnamefont {S.}~\bibnamefont {Haroche}},\
  }\bibfield  {title} {\bibinfo {title} {Progressive field-state collapse and
  quantum non-demolition photon counting},\ }\bibfield  {journal} {\bibinfo
  {journal} {Nature}\ }\textbf {\bibinfo {volume} {448}},\ \href
  {https://doi.org/10.1038/nature06057} {10.1038/nature06057} (\bibinfo {year}
  {2007})\BibitemShut {NoStop}%
\bibitem [{\citenamefont {Sun}\ \emph {et~al.}(2014)\citenamefont {Sun},
  \citenamefont {Petrenko}, \citenamefont {Leghtas}, \citenamefont {Vlastakis},
  \citenamefont {Kirchmair}, \citenamefont {Sliwa}, \citenamefont {Narla},
  \citenamefont {Hatridge}, \citenamefont {Shankar}, \citenamefont {Blumoff},
  \citenamefont {Frunzio}, \citenamefont {Mirrahimi}, \citenamefont {Devoret},\
  and\ \citenamefont {Schoelkopf}}]{sun_tracking_2014}%
  \BibitemOpen
  \bibfield  {author} {\bibinfo {author} {\bibfnamefont {L.}~\bibnamefont
  {Sun}}, \bibinfo {author} {\bibfnamefont {A.}~\bibnamefont {Petrenko}},
  \bibinfo {author} {\bibfnamefont {Z.}~\bibnamefont {Leghtas}}, \bibinfo
  {author} {\bibfnamefont {B.}~\bibnamefont {Vlastakis}}, \bibinfo {author}
  {\bibfnamefont {G.}~\bibnamefont {Kirchmair}}, \bibinfo {author}
  {\bibfnamefont {K.~M.}\ \bibnamefont {Sliwa}}, \bibinfo {author}
  {\bibfnamefont {A.}~\bibnamefont {Narla}}, \bibinfo {author} {\bibfnamefont
  {M.}~\bibnamefont {Hatridge}}, \bibinfo {author} {\bibfnamefont
  {S.}~\bibnamefont {Shankar}}, \bibinfo {author} {\bibfnamefont
  {J.}~\bibnamefont {Blumoff}}, \bibinfo {author} {\bibfnamefont
  {L.}~\bibnamefont {Frunzio}}, \bibinfo {author} {\bibfnamefont
  {M.}~\bibnamefont {Mirrahimi}}, \bibinfo {author} {\bibfnamefont {M.~H.}\
  \bibnamefont {Devoret}},\ and\ \bibinfo {author} {\bibfnamefont {R.~J.}\
  \bibnamefont {Schoelkopf}},\ }\bibfield  {title} {\bibinfo {title} {Tracking
  photon jumps with repeated quantum non-demolition parity measurements},\
  }\bibfield  {journal} {\bibinfo  {journal} {Nature}\ }\textbf {\bibinfo
  {volume} {511}},\ \href {https://doi.org/10.1038/nature13436}
  {10.1038/nature13436} (\bibinfo {year} {2014})\BibitemShut {NoStop}%
\bibitem [{\citenamefont {Essig}\ \emph {et~al.}(2020)\citenamefont {Essig},
  \citenamefont {Ficheux}, \citenamefont {Peronnin}, \citenamefont {Cottet},
  \citenamefont {Lescanne}, \citenamefont {Sarlette}, \citenamefont {Rouchon},
  \citenamefont {Leghtas},\ and\ \citenamefont
  {Huard}}]{essig_multiplexed_2020}%
  \BibitemOpen
  \bibfield  {author} {\bibinfo {author} {\bibfnamefont {A.}~\bibnamefont
  {Essig}}, \bibinfo {author} {\bibfnamefont {Q.}~\bibnamefont {Ficheux}},
  \bibinfo {author} {\bibfnamefont {T.}~\bibnamefont {Peronnin}}, \bibinfo
  {author} {\bibfnamefont {N.}~\bibnamefont {Cottet}}, \bibinfo {author}
  {\bibfnamefont {R.}~\bibnamefont {Lescanne}}, \bibinfo {author}
  {\bibfnamefont {A.}~\bibnamefont {Sarlette}}, \bibinfo {author}
  {\bibfnamefont {P.}~\bibnamefont {Rouchon}}, \bibinfo {author} {\bibfnamefont
  {Z.}~\bibnamefont {Leghtas}},\ and\ \bibinfo {author} {\bibfnamefont
  {B.}~\bibnamefont {Huard}},\ }\bibfield  {title} {\bibinfo {title}
  {Multiplexed photon number measurement},\ }\href
  {http://arxiv.org/abs/2001.03217} {\bibfield  {journal} {\bibinfo  {journal}
  {arXiv:2001.03217 [cond-mat, physics:quant-ph]}\ } (\bibinfo {year}
  {2020})}\BibitemShut {NoStop}%
\bibitem [{\citenamefont {Dassonneville}\ \emph {et~al.}(2020)\citenamefont
  {Dassonneville}, \citenamefont {Assouly}, \citenamefont {Peronnin},
  \citenamefont {Rouchon},\ and\ \citenamefont
  {Huard}}]{dassonneville_number-resolved_2020}%
  \BibitemOpen
  \bibfield  {author} {\bibinfo {author} {\bibfnamefont {R.}~\bibnamefont
  {Dassonneville}}, \bibinfo {author} {\bibfnamefont {R.}~\bibnamefont
  {Assouly}}, \bibinfo {author} {\bibfnamefont {T.}~\bibnamefont {Peronnin}},
  \bibinfo {author} {\bibfnamefont {P.}~\bibnamefont {Rouchon}},\ and\ \bibinfo
  {author} {\bibfnamefont {B.}~\bibnamefont {Huard}},\ }\bibfield  {title}
  {\bibinfo {title} {Number-{Resolved} {Photocounter} for {Propagating}
  {Microwave} {Mode}},\ }\href
  {https://doi.org/10.1103/PhysRevApplied.14.044022} {\bibfield  {journal}
  {\bibinfo  {journal} {Physical Review Applied}\ }\textbf {\bibinfo {volume}
  {14}},\ \bibinfo {pages} {044022} (\bibinfo {year} {2020})},\ \bibinfo {note}
  {publisher: American Physical Society}\BibitemShut {NoStop}%
\bibitem [{\citenamefont {Wang}\ \emph {et~al.}(2020)\citenamefont {Wang},
  \citenamefont {Curtis}, \citenamefont {Lester}, \citenamefont {Zhang},
  \citenamefont {Gao}, \citenamefont {Freeze}, \citenamefont {Batista},
  \citenamefont {Vaccaro}, \citenamefont {Chuang}, \citenamefont {Frunzio},
  \citenamefont {Jiang}, \citenamefont {Girvin},\ and\ \citenamefont
  {Schoelkopf}}]{wang_efficient_2020}%
  \BibitemOpen
  \bibfield  {author} {\bibinfo {author} {\bibfnamefont {C.~S.}\ \bibnamefont
  {Wang}}, \bibinfo {author} {\bibfnamefont {J.~C.}\ \bibnamefont {Curtis}},
  \bibinfo {author} {\bibfnamefont {B.~J.}\ \bibnamefont {Lester}}, \bibinfo
  {author} {\bibfnamefont {Y.}~\bibnamefont {Zhang}}, \bibinfo {author}
  {\bibfnamefont {Y.~Y.}\ \bibnamefont {Gao}}, \bibinfo {author} {\bibfnamefont
  {J.}~\bibnamefont {Freeze}}, \bibinfo {author} {\bibfnamefont {V.~S.}\
  \bibnamefont {Batista}}, \bibinfo {author} {\bibfnamefont {P.~H.}\
  \bibnamefont {Vaccaro}}, \bibinfo {author} {\bibfnamefont {I.~L.}\
  \bibnamefont {Chuang}}, \bibinfo {author} {\bibfnamefont {L.}~\bibnamefont
  {Frunzio}}, \bibinfo {author} {\bibfnamefont {L.}~\bibnamefont {Jiang}},
  \bibinfo {author} {\bibfnamefont {S.~M.}\ \bibnamefont {Girvin}},\ and\
  \bibinfo {author} {\bibfnamefont {R.~J.}\ \bibnamefont {Schoelkopf}},\
  }\bibfield  {title} {\bibinfo {title} {Efficient {Multiphoton} {Sampling} of
  {Molecular} {Vibronic} {Spectra} on a {Superconducting} {Bosonic}
  {Processor}},\ }\href {https://doi.org/10.1103/PhysRevX.10.021060} {\bibfield
   {journal} {\bibinfo  {journal} {Phys. Rev. X}\ }\textbf {\bibinfo {volume}
  {10}},\ \bibinfo {pages} {021060} (\bibinfo {year} {2020})}\BibitemShut
  {NoStop}%
\bibitem [{\citenamefont {Haroche}\ and\ \citenamefont
  {Raimond}()}]{haroche_exploring_nodate}%
  \BibitemOpen
  \bibfield  {author} {\bibinfo {author} {\bibfnamefont {S.}~\bibnamefont
  {Haroche}}\ and\ \bibinfo {author} {\bibfnamefont {J.-M.}\ \bibnamefont
  {Raimond}},\ }\href
  {https://oxford.universitypressscholarship.com/view/10.1093/acprof:oso/9780198509141.001.0001/acprof-9780198509141}
  {\emph {\bibinfo {title} {Exploring the {Quantum}: {Atoms}, {Cavities}, and
  {Photons}}}}\ (\bibinfo  {publisher} {Oxford University Press})\BibitemShut
  {NoStop}%
\bibitem [{\citenamefont {Elder}\ \emph {et~al.}(2020)\citenamefont {Elder},
  \citenamefont {Wang}, \citenamefont {Reinhold}, \citenamefont {Hann},
  \citenamefont {Chou}, \citenamefont {Lester}, \citenamefont {Rosenblum},
  \citenamefont {Frunzio}, \citenamefont {Jiang},\ and\ \citenamefont
  {Schoelkopf}}]{elder_high-fidelity_2020}%
  \BibitemOpen
  \bibfield  {author} {\bibinfo {author} {\bibfnamefont {S.~S.}\ \bibnamefont
  {Elder}}, \bibinfo {author} {\bibfnamefont {C.~S.}\ \bibnamefont {Wang}},
  \bibinfo {author} {\bibfnamefont {P.}~\bibnamefont {Reinhold}}, \bibinfo
  {author} {\bibfnamefont {C.~T.}\ \bibnamefont {Hann}}, \bibinfo {author}
  {\bibfnamefont {K.~S.}\ \bibnamefont {Chou}}, \bibinfo {author}
  {\bibfnamefont {B.~J.}\ \bibnamefont {Lester}}, \bibinfo {author}
  {\bibfnamefont {S.}~\bibnamefont {Rosenblum}}, \bibinfo {author}
  {\bibfnamefont {L.}~\bibnamefont {Frunzio}}, \bibinfo {author} {\bibfnamefont
  {L.}~\bibnamefont {Jiang}},\ and\ \bibinfo {author} {\bibfnamefont {R.~J.}\
  \bibnamefont {Schoelkopf}},\ }\bibfield  {title} {\bibinfo {title}
  {High-{Fidelity} {Measurement} of {Qubits} {Encoded} in {Multilevel}
  {Superconducting} {Circuits}},\ }\href
  {https://doi.org/10.1103/PhysRevX.10.011001} {\bibfield  {journal} {\bibinfo
  {journal} {Phys. Rev. X}\ }\textbf {\bibinfo {volume} {10}},\ \bibinfo
  {pages} {011001} (\bibinfo {year} {2020})}\BibitemShut {NoStop}%
\bibitem [{\citenamefont {Reinhold}\ \emph {et~al.}(2020)\citenamefont
  {Reinhold}, \citenamefont {Rosenblum}, \citenamefont {Ma}, \citenamefont
  {Frunzio}, \citenamefont {Jiang},\ and\ \citenamefont
  {Schoelkopf}}]{reinhold_error-corrected_2020}%
  \BibitemOpen
  \bibfield  {author} {\bibinfo {author} {\bibfnamefont {P.}~\bibnamefont
  {Reinhold}}, \bibinfo {author} {\bibfnamefont {S.}~\bibnamefont {Rosenblum}},
  \bibinfo {author} {\bibfnamefont {W.-L.}\ \bibnamefont {Ma}}, \bibinfo
  {author} {\bibfnamefont {L.}~\bibnamefont {Frunzio}}, \bibinfo {author}
  {\bibfnamefont {L.}~\bibnamefont {Jiang}},\ and\ \bibinfo {author}
  {\bibfnamefont {R.~J.}\ \bibnamefont {Schoelkopf}},\ }\bibfield  {title}
  {\bibinfo {title} {Error-corrected gates on an encoded qubit},\ }\bibfield
  {journal} {\bibinfo  {journal} {Nature Physics}\ }\textbf {\bibinfo {volume}
  {16}},\ \href {https://doi.org/10.1038/s41567-020-0931-8}
  {10.1038/s41567-020-0931-8} (\bibinfo {year} {2020})\BibitemShut {NoStop}%
\bibitem [{\citenamefont {Temme}\ \emph {et~al.}(2017)\citenamefont {Temme},
  \citenamefont {Bravyi},\ and\ \citenamefont {Gambetta}}]{temme_error_2017}%
  \BibitemOpen
  \bibfield  {author} {\bibinfo {author} {\bibfnamefont {K.}~\bibnamefont
  {Temme}}, \bibinfo {author} {\bibfnamefont {S.}~\bibnamefont {Bravyi}},\ and\
  \bibinfo {author} {\bibfnamefont {J.~M.}\ \bibnamefont {Gambetta}},\
  }\bibfield  {title} {\bibinfo {title} {Error {Mitigation} for {Short}-{Depth}
  {Quantum} {Circuits}},\ }\href
  {https://doi.org/10.1103/PhysRevLett.119.180509} {\bibfield  {journal}
  {\bibinfo  {journal} {Phys. Rev. Lett.}\ }\textbf {\bibinfo {volume} {119}},\
  \bibinfo {pages} {180509} (\bibinfo {year} {2017})}\BibitemShut {NoStop}%
\bibitem [{\citenamefont {Li}\ and\ \citenamefont
  {Benjamin}(2017)}]{li_efficient_2017}%
  \BibitemOpen
  \bibfield  {author} {\bibinfo {author} {\bibfnamefont {Y.}~\bibnamefont
  {Li}}\ and\ \bibinfo {author} {\bibfnamefont {S.~C.}\ \bibnamefont
  {Benjamin}},\ }\bibfield  {title} {\bibinfo {title} {Efficient {Variational}
  {Quantum} {Simulator} {Incorporating} {Active} {Error} {Minimization}},\
  }\href {https://doi.org/10.1103/PhysRevX.7.021050} {\bibfield  {journal}
  {\bibinfo  {journal} {Phys. Rev. X}\ }\textbf {\bibinfo {volume} {7}},\
  \bibinfo {pages} {021050} (\bibinfo {year} {2017})}\BibitemShut {NoStop}%
\bibitem [{\citenamefont {Kandala}\ \emph {et~al.}(2019)\citenamefont
  {Kandala}, \citenamefont {Temme}, \citenamefont {Córcoles}, \citenamefont
  {Mezzacapo}, \citenamefont {Chow},\ and\ \citenamefont
  {Gambetta}}]{kandala_error_2019}%
  \BibitemOpen
  \bibfield  {author} {\bibinfo {author} {\bibfnamefont {A.}~\bibnamefont
  {Kandala}}, \bibinfo {author} {\bibfnamefont {K.}~\bibnamefont {Temme}},
  \bibinfo {author} {\bibfnamefont {A.~D.}\ \bibnamefont {Córcoles}}, \bibinfo
  {author} {\bibfnamefont {A.}~\bibnamefont {Mezzacapo}}, \bibinfo {author}
  {\bibfnamefont {J.~M.}\ \bibnamefont {Chow}},\ and\ \bibinfo {author}
  {\bibfnamefont {J.~M.}\ \bibnamefont {Gambetta}},\ }\bibfield  {title}
  {\bibinfo {title} {Error mitigation extends the computational reach of a
  noisy quantum processor},\ }\bibfield  {journal} {\bibinfo  {journal}
  {Nature}\ }\textbf {\bibinfo {volume} {567}},\ \href
  {https://doi.org/10.1038/s41586-019-1040-7} {10.1038/s41586-019-1040-7}
  (\bibinfo {year} {2019})\BibitemShut {NoStop}%
\bibitem [{\citenamefont {Nielsen}\ and\ \citenamefont
  {Chuang}(2011)}]{nielsen_quantum_2011}%
  \BibitemOpen
  \bibfield  {author} {\bibinfo {author} {\bibfnamefont {M.~A.}\ \bibnamefont
  {Nielsen}}\ and\ \bibinfo {author} {\bibfnamefont {I.~L.}\ \bibnamefont
  {Chuang}},\ }\href@noop {} {\emph {\bibinfo {title} {Quantum {Computation}
  and {Quantum} {Information}: 10th {Anniversary} {Edition}}}},\ \bibinfo
  {edition} {10th}\ ed.\ (\bibinfo  {publisher} {Cambridge University Press},\
  \bibinfo {address} {USA},\ \bibinfo {year} {2011})\BibitemShut {NoStop}%
\bibitem [{\citenamefont {Maciejewski}\ \emph {et~al.}(2020)\citenamefont
  {Maciejewski}, \citenamefont {Zimborás},\ and\ \citenamefont
  {Oszmaniec}}]{maciejewski_mitigation_2020}%
  \BibitemOpen
  \bibfield  {author} {\bibinfo {author} {\bibfnamefont {F.~B.}\ \bibnamefont
  {Maciejewski}}, \bibinfo {author} {\bibfnamefont {Z.}~\bibnamefont
  {Zimborás}},\ and\ \bibinfo {author} {\bibfnamefont {M.}~\bibnamefont
  {Oszmaniec}},\ }\bibfield  {title} {\bibinfo {title} {Mitigation of readout
  noise in near-term quantum devices by classical post-processing based on
  detector tomography},\ }\href {https://doi.org/10.22331/q-2020-04-24-257}
  {\bibfield  {journal} {\bibinfo  {journal} {Quantum}\ }\textbf {\bibinfo
  {volume} {4}},\ \bibinfo {pages} {257} (\bibinfo {year} {2020})}\BibitemShut
  {NoStop}%
\bibitem [{\citenamefont {Dreau}\ \emph {et~al.}(2013)\citenamefont {Dreau},
  \citenamefont {Spinicelli}, \citenamefont {Maze}, \citenamefont {Roch},\ and\
  \citenamefont {Jacques}}]{dreau_single-shot_2013}%
  \BibitemOpen
  \bibfield  {author} {\bibinfo {author} {\bibfnamefont {A.}~\bibnamefont
  {Dreau}}, \bibinfo {author} {\bibfnamefont {P.}~\bibnamefont {Spinicelli}},
  \bibinfo {author} {\bibfnamefont {J.~R.}\ \bibnamefont {Maze}}, \bibinfo
  {author} {\bibfnamefont {J.-F.}\ \bibnamefont {Roch}},\ and\ \bibinfo
  {author} {\bibfnamefont {V.}~\bibnamefont {Jacques}},\ }\bibfield  {title}
  {\bibinfo {title} {Single-{Shot} {Readout} of {Multiple} {Nuclear} {Spin}
  {Qubits} in {Diamond} under {Ambient} {Conditions}},\ }\href
  {https://doi.org/10.1103/PhysRevLett.110.060502} {\bibfield  {journal}
  {\bibinfo  {journal} {Phys. Rev. Lett.}\ }\textbf {\bibinfo {volume} {110}},\
  \bibinfo {pages} {060502} (\bibinfo {year} {2013})}\BibitemShut {NoStop}%
\bibitem [{\citenamefont {Gammelmark}\ \emph {et~al.}(2013)\citenamefont
  {Gammelmark}, \citenamefont {Julsgaard},\ and\ \citenamefont
  {Molmer}}]{gammelmark_past_2013}%
  \BibitemOpen
  \bibfield  {author} {\bibinfo {author} {\bibfnamefont {S.}~\bibnamefont
  {Gammelmark}}, \bibinfo {author} {\bibfnamefont {B.}~\bibnamefont
  {Julsgaard}},\ and\ \bibinfo {author} {\bibfnamefont {K.}~\bibnamefont
  {Molmer}},\ }\bibfield  {title} {\bibinfo {title} {Past {Quantum} {States} of
  a {Monitored} {System}},\ }\href
  {https://doi.org/10.1103/PhysRevLett.111.160401} {\bibfield  {journal}
  {\bibinfo  {journal} {Phys. Rev. Lett.}\ }\textbf {\bibinfo {volume} {111}},\
  \bibinfo {pages} {160401} (\bibinfo {year} {2013})}\BibitemShut {NoStop}%
\bibitem [{\citenamefont {Ng}\ and\ \citenamefont
  {Tsang}(2014)}]{ng_optimal_2014}%
  \BibitemOpen
  \bibfield  {author} {\bibinfo {author} {\bibfnamefont {S.}~\bibnamefont
  {Ng}}\ and\ \bibinfo {author} {\bibfnamefont {M.}~\bibnamefont {Tsang}},\
  }\bibfield  {title} {\bibinfo {title} {Optimal signal processing for
  continuous qubit readout},\ }\href
  {https://doi.org/10.1103/PhysRevA.90.022325} {\bibfield  {journal} {\bibinfo
  {journal} {Phys. Rev. A}\ }\textbf {\bibinfo {volume} {90}},\ \bibinfo
  {pages} {022325} (\bibinfo {year} {2014})}\BibitemShut {NoStop}%
\bibitem [{\citenamefont {Wölk}\ \emph {et~al.}(2015)\citenamefont {Wölk},
  \citenamefont {Piltz}, \citenamefont {Sriarunothai},\ and\ \citenamefont
  {Wunderlich}}]{wolk_state_2015}%
  \BibitemOpen
  \bibfield  {author} {\bibinfo {author} {\bibfnamefont {S.}~\bibnamefont
  {Wölk}}, \bibinfo {author} {\bibfnamefont {C.}~\bibnamefont {Piltz}},
  \bibinfo {author} {\bibfnamefont {T.}~\bibnamefont {Sriarunothai}},\ and\
  \bibinfo {author} {\bibfnamefont {C.}~\bibnamefont {Wunderlich}},\ }\bibfield
   {title} {\bibinfo {title} {State selective detection of hyperfine qubits},\
  }\href {https://doi.org/10.1088/0953-4075/48/7/075101} {\bibfield  {journal}
  {\bibinfo  {journal} {J. Phys. B: At. Mol. Opt. Phys.}\ }\textbf {\bibinfo
  {volume} {48}},\ \bibinfo {pages} {075101} (\bibinfo {year}
  {2015})}\BibitemShut {NoStop}%
\bibitem [{\citenamefont {Martinez}\ \emph {et~al.}(2020)\citenamefont
  {Martinez}, \citenamefont {Rosen},\ and\ \citenamefont
  {DuBois}}]{martinez2020improving}%
  \BibitemOpen
  \bibfield  {author} {\bibinfo {author} {\bibfnamefont {L.~A.}\ \bibnamefont
  {Martinez}}, \bibinfo {author} {\bibfnamefont {Y.~J.}\ \bibnamefont
  {Rosen}},\ and\ \bibinfo {author} {\bibfnamefont {J.~L.}\ \bibnamefont
  {DuBois}},\ }\bibfield  {title} {\bibinfo {title} {Improving qubit readout
  with hidden markov models},\ }\href@noop {} {\  (\bibinfo {year} {2020})},\
  \Eprint {https://arxiv.org/abs/2006.00109} {arXiv:2006.00109 [quant-ph]}
  \BibitemShut {NoStop}%
\bibitem [{\citenamefont {Hann}\ \emph {et~al.}(2018)\citenamefont {Hann},
  \citenamefont {Elder}, \citenamefont {Wang}, \citenamefont {Chou},
  \citenamefont {Schoelkopf},\ and\ \citenamefont {Jiang}}]{hann_robust_2018}%
  \BibitemOpen
  \bibfield  {author} {\bibinfo {author} {\bibfnamefont {C.~T.}\ \bibnamefont
  {Hann}}, \bibinfo {author} {\bibfnamefont {S.~S.}\ \bibnamefont {Elder}},
  \bibinfo {author} {\bibfnamefont {C.~S.}\ \bibnamefont {Wang}}, \bibinfo
  {author} {\bibfnamefont {K.}~\bibnamefont {Chou}}, \bibinfo {author}
  {\bibfnamefont {R.~J.}\ \bibnamefont {Schoelkopf}},\ and\ \bibinfo {author}
  {\bibfnamefont {L.}~\bibnamefont {Jiang}},\ }\bibfield  {title} {\bibinfo
  {title} {Robust readout of bosonic qubits in the dispersive coupling
  regime},\ }\href {https://doi.org/10.1103/PhysRevA.98.022305} {\bibfield
  {journal} {\bibinfo  {journal} {Phys. Rev. A}\ }\textbf {\bibinfo {volume}
  {98}},\ \bibinfo {pages} {022305} (\bibinfo {year} {2018})}\BibitemShut
  {NoStop}%
\bibitem [{\citenamefont {Gasparinetti}\ \emph {et~al.}(2016)\citenamefont
  {Gasparinetti}, \citenamefont {Berger}, \citenamefont {Abdumalikov},
  \citenamefont {Pechal}, \citenamefont {Filipp},\ and\ \citenamefont
  {Wallraff}}]{gasparinetti_measurement_2016}%
  \BibitemOpen
  \bibfield  {author} {\bibinfo {author} {\bibfnamefont {S.}~\bibnamefont
  {Gasparinetti}}, \bibinfo {author} {\bibfnamefont {S.}~\bibnamefont
  {Berger}}, \bibinfo {author} {\bibfnamefont {A.~A.}\ \bibnamefont
  {Abdumalikov}}, \bibinfo {author} {\bibfnamefont {M.}~\bibnamefont {Pechal}},
  \bibinfo {author} {\bibfnamefont {S.}~\bibnamefont {Filipp}},\ and\ \bibinfo
  {author} {\bibfnamefont {A.~J.}\ \bibnamefont {Wallraff}},\ }\bibfield
  {title} {\bibinfo {title} {Measurement of a vacuum-induced geometric phase},\
  }\href {https://doi.org/10.1126/sciadv.1501732} {\bibfield  {journal}
  {\bibinfo  {journal} {Science Advances}\ }\textbf {\bibinfo {volume} {2}},\
  \bibinfo {pages} {e1501732} (\bibinfo {year} {2016})}\BibitemShut {NoStop}%
\bibitem [{\citenamefont {Lundeen}\ \emph {et~al.}(2009)\citenamefont
  {Lundeen}, \citenamefont {Feito}, \citenamefont {Coldenstrodt-Ronge},
  \citenamefont {Pregnell}, \citenamefont {Silberhorn}, \citenamefont {Ralph},
  \citenamefont {Eisert}, \citenamefont {Plenio},\ and\ \citenamefont
  {Walmsley}}]{Lundeen2009}%
  \BibitemOpen
  \bibfield  {author} {\bibinfo {author} {\bibfnamefont {J.~S.}\ \bibnamefont
  {Lundeen}}, \bibinfo {author} {\bibfnamefont {A.}~\bibnamefont {Feito}},
  \bibinfo {author} {\bibfnamefont {H.}~\bibnamefont {Coldenstrodt-Ronge}},
  \bibinfo {author} {\bibfnamefont {K.~L.}\ \bibnamefont {Pregnell}}, \bibinfo
  {author} {\bibfnamefont {C.}~\bibnamefont {Silberhorn}}, \bibinfo {author}
  {\bibfnamefont {T.~C.}\ \bibnamefont {Ralph}}, \bibinfo {author}
  {\bibfnamefont {J.}~\bibnamefont {Eisert}}, \bibinfo {author} {\bibfnamefont
  {M.~B.}\ \bibnamefont {Plenio}},\ and\ \bibinfo {author} {\bibfnamefont
  {I.~A.}\ \bibnamefont {Walmsley}},\ }\bibfield  {title} {\bibinfo {title}
  {Tomography of quantum detectors},\ }\href
  {https://doi.org/10.1038/nphys1133} {\bibfield  {journal} {\bibinfo
  {journal} {Nature Physics}\ }\textbf {\bibinfo {volume} {5}},\ \bibinfo
  {pages} {27} (\bibinfo {year} {2009})}\BibitemShut {NoStop}%
\bibitem [{\citenamefont {Khaneja}\ \emph {et~al.}(2005)\citenamefont
  {Khaneja}, \citenamefont {Reiss}, \citenamefont {Kehlet}, \citenamefont
  {Schulte-Herbrüggen},\ and\ \citenamefont
  {Glaser}}]{khaneja_optimal_control}%
  \BibitemOpen
  \bibfield  {author} {\bibinfo {author} {\bibfnamefont {N.}~\bibnamefont
  {Khaneja}}, \bibinfo {author} {\bibfnamefont {T.}~\bibnamefont {Reiss}},
  \bibinfo {author} {\bibfnamefont {C.}~\bibnamefont {Kehlet}}, \bibinfo
  {author} {\bibfnamefont {T.}~\bibnamefont {Schulte-Herbrüggen}},\ and\
  \bibinfo {author} {\bibfnamefont {S.~J.}\ \bibnamefont {Glaser}},\ }\bibfield
   {title} {\bibinfo {title} {Optimal control of coupled spin dynamics: design
  of nmr pulse sequences by gradient ascent algorithms},\ }\href
  {https://doi.org/https://doi.org/10.1016/j.jmr.2004.11.004} {\bibfield
  {journal} {\bibinfo  {journal} {Journal of Magnetic Resonance}\ }\textbf
  {\bibinfo {volume} {172}},\ \bibinfo {pages} {296 } (\bibinfo {year}
  {2005})}\BibitemShut {NoStop}%
\bibitem [{\citenamefont {{de Fouquieres}}\ \emph {et~al.}(2011)\citenamefont
  {{de Fouquieres}}, \citenamefont {Schirmer}, \citenamefont {Glaser},\ and\
  \citenamefont {Kuprov}}]{deFouquieres_2nd_order_GRAPE}%
  \BibitemOpen
  \bibfield  {author} {\bibinfo {author} {\bibfnamefont {P.}~\bibnamefont {{de
  Fouquieres}}}, \bibinfo {author} {\bibfnamefont {S.}~\bibnamefont
  {Schirmer}}, \bibinfo {author} {\bibfnamefont {S.}~\bibnamefont {Glaser}},\
  and\ \bibinfo {author} {\bibfnamefont {I.}~\bibnamefont {Kuprov}},\
  }\bibfield  {title} {\bibinfo {title} {Second order gradient ascent pulse
  engineering},\ }\href
  {https://doi.org/https://doi.org/10.1016/j.jmr.2011.07.023} {\bibfield
  {journal} {\bibinfo  {journal} {Journal of Magnetic Resonance}\ }\textbf
  {\bibinfo {volume} {212}},\ \bibinfo {pages} {412 } (\bibinfo {year}
  {2011})}\BibitemShut {NoStop}%
\bibitem [{\citenamefont {{Shannon}}(1948)}]{shannon_entropy}%
  \BibitemOpen
  \bibfield  {author} {\bibinfo {author} {\bibfnamefont {C.~E.}\ \bibnamefont
  {{Shannon}}},\ }\bibfield  {title} {\bibinfo {title} {A mathematical theory
  of communication},\ }\href
  {https://doi.org/10.1002/j.1538-7305.1948.tb01338.x} {\bibfield  {journal}
  {\bibinfo  {journal} {The Bell System Technical Journal}\ }\textbf {\bibinfo
  {volume} {27}},\ \bibinfo {pages} {379} (\bibinfo {year} {1948})}\BibitemShut
  {NoStop}%
\end{thebibliography}%

\end{document}